\documentclass[final,5p,times,twocolumn]{elsarticle}

\usepackage{amssymb}
\usepackage{lineno}

\usepackage{amsmath}
\usepackage{amssymb}
\usepackage{hyperref}

\newcommand{\Figure}[1]{Figure~\ref{fig:#1}}

\newcommand{\Table}[1]{Table~\ref{tab:#1}}

\newcommand{\Eq}[1]{Eq.~\ref{eq:#1}}

\newcommand{\Section}[1]{Section~\ref{sec:#1}}

\usepackage{bbold}
\renewcommand\vec{\mathbf}
\newcommand{\mat}{\mathbf}

\DeclareMathOperator*{\argmin}{argmin}

\DeclareMathOperator*{\diag}{diag}

\usepackage{algorithm}% http://ctan.org/pkg/algorithm
\usepackage{algpseudocode}% http://ctan.org/pkg/algorithmicx

\usepackage{enumitem}

\usepackage{caption}
\usepackage{subcaption}

\usepackage{multirow}
\usepackage{graphicx}

\usepackage{sidecap}

\usepackage{float}

\makeatletter
\newcommand\xlabel[2][]{\phantomsection\def\@currentlabelname{#1}\label{#2}}
\makeatother

\usepackage{geometry}

\usepackage[flushleft]{threeparttable}

\begin{document}

\begin{frontmatter}

\title{On Retrospective $k$-space Subsampling schemes For Deep MRI Reconstruction}

\author[a]{George Yiasemis}
% \ead{g.yiasemis@nki.nl}
% \ead[url]{https://aiforoncology.nl/}
% \cortext[cor1]{Corresponding Author}

\author[b]{Clara I. Sánchez}
\author[a]{Jan-Jakob Sonke}
\author[a,b,c]{Jonas Teuwen}
\affiliation[a]{organization={AI for Oncology, Netherlands Cancer Institute},%Department and Organization
            addressline={Plesmanlaan 121}, 
            city={Amsterdam},
            postcode={1066 CX}, 
            country={Netherlands}}

\affiliation[b]{organization={University of Amsterdam},%Department and Organization
            addressline={Science Park 904}, 
            city={Amsterdam},
            postcode={1098 XH}, 
            country={Netherlands}}

\affiliation[c]{organization={Radboud University Medical Center, Department of Medical Imaging},%Department and Organization
            addressline={Geert Grooteplein Zuid 10 }, 
            city={Nijmegen},
            postcode={6525 GA}, 
            country={Netherlands}}

\begin{abstract}
% Purpose 
Acquiring fully-sampled MRI $k$-space data is time-consuming, and collecting accelerated data can reduce the acquisition time. Employing 2D Cartesian-rectilinear subsampling schemes is a conventional approach for accelerated acquisitions; however, this often results in imprecise reconstructions, even with the use of Deep Learning (DL), especially at high acceleration factors. Non-rectilinear or non-Cartesian trajectories can be implemented in MRI scanners as alternative subsampling options. This work investigates the impact of the $k$-space subsampling scheme on the quality of reconstructed accelerated MRI measurements produced by trained DL models. The Recurrent Variational Network (RecurrentVarNet) was used as the DL-based MRI-reconstruction architecture. Cartesian, fully-sampled multi-coil $k$-space measurements from three datasets were retrospectively subsampled with different accelerations using eight distinct subsampling schemes: four Cartesian-rectilinear, two Cartesian non-rectilinear, and two non-Cartesian. Experiments were conducted in two frameworks: scheme-specific, where a distinct model was trained and evaluated for each dataset-subsampling scheme pair, and multi-scheme, where for each dataset a single model was trained on data randomly subsampled by any of the eight schemes and evaluated on data subsampled by all schemes. In both frameworks, RecurrentVarNets trained and evaluated on non-rectilinearly subsampled data demonstrated superior performance, particularly for high accelerations. In the multi-scheme setting, reconstruction performance on rectilinearly subsampled data improved when compared to the scheme-specific experiments. Our findings demonstrate the potential for using DL-based methods, trained on non-rectilinearly subsampled measurements, to optimize scan time and image quality.
\end{abstract}


% \begin{highlights}
% \item The study compares various 2D retrospective k-space subsampling patterns and their effect on the quality of deep learning-based MRI reconstructions.
% \item Results show that non-rectilinear Cartesian and non-Cartesian subsampling patterns result in higher-quality reconstructions compared to conventional rectilinear patterns.
% % \item The study suggests that future studies should explore repeating the experiments using prospectively subsampled data and incorporating non-Cartesian data into deep learning pipelines.
% \end{highlights}

\begin{keyword}
Deep MRI Reconstruction \sep Retrospective $k$-space Subsampling \sep Non-rectilinear Subsampling \sep Non-Cartesian Subsampling \sep Recurrent Variational Network% %% PACS codes here, in the form: \PACS code \sep code
% %% MSC codes here, in the form: \MSC code \sep code
% %% or \MSC[2008] code \sep code (2000 is the default)
\end{keyword}

\end{frontmatter}

% \linenumbers


\section{Introduction}
\label{sect:intro}

Magnetic Resonance Imaging (MRI) is a widely used medical imaging modality known for its ability to produce high soft tissue contrast, high-resolution images without the use of ionizing radiation. Its applications range from diagnosis and treatment planning to real-time MR-guided tasks such as surgery or radiotherapy.  However, the long acquisition times of the MRI measurements, known as $k$-space, hinder the full potential of MRI-guidance. Reducing these times could make MRI-guidance cheaper and extend its functionality in real-time settings.

Over the past two decades, several methods have been put to use in clinical practice for accelerating the MRI acquisition. The two most conventionally applied methods to-date are  Parallel Imaging (PI) and Compressed Sensing (CS), which are usually both incorporated in modern state-of-the-art MRI scanners. 

Compressed Sensing aims to reconstruct images from incoherently subsampled $k$-space measurements through mathematical optimization techniques \cite{1614066, cpa.20124, cs-1580791, 21391}. Subsampling the $k$-space is, in general, a violation of the Nyquist-Shannon sampling criterion \cite{1697831} and thus prone to producing aliasing artifacts. CS reconstruction algorithms employ optimization methods like Total Variation (TV) \cite{4587391} to obtain faithful images from sparse input signals.

Parallel Imaging on the other hand, employs an array of multiple - instead of one - radio-frequency receiver coils which measure sets of spatially localised $k$-space frequencies while maintaining the same spatial resolution \cite{Griswold_2002, parallel-https://doi.org/10.1002/nbm.1042, Lustig_2010}. Each independent receiver coil receives distinct measurements corresponding to their spatial location in relation to the scanned object.

With the recent advancements in Deep Learning (DL) and Computer Vision (CV), a plethora of algorithms have emerged targeting to solve imaging inverse problems, with Accelerated MRI Reconstruction being a par excellence example.  By utilizing CS optimization approaches and PI, numerous DL-based methods involving convolutional neural networks (CNNs) have been proposed in the literature \cite{https://doi.org/10.1002/mrm.26977, Eo_2018, LONNING201964, Sriram_2020_CVPR, Souza_2020, Yiasemis_2022_CVPR} applied to the task of Accelerated MRI Reconstruction. These methods are usually trained in a supervised manner using retrospectively subsampled (from available fully-sampled) $k$-space datasets and their target is to make a prediction of the fully-sampled $k$-space or its image reconstruction. 

Rectilinear Cartesian patterns constitute the most commonly applied subsampling techniques in clinical 2D MRI acquisitions. Subsequently, DL-based Accelerated MRI Reconstruction applications utilize rectilinear subsampling masks to retrospectively subsample fully-sampled data. However, a variety of prospective and retrospective non-rectilinear Cartesian sampling and subsampling patterns exist. For instance, non-Cartesian patterns such as radial or spiral are being applied in real-time MRI acquisitions due to the fact that they are less susceptible to motion compared to Cartesian ones \cite{Feng2014-vk}. The authors in \citenum{Yiasemis_2022_radial} by employing a deep neural network architecture, namely the Recurrent Inference Machine (RIM) \cite{LONNING201964}, explored the effects of training RIMs by applying either rectilinear or radial retrospective subsampling, and concluded that the RIM trained using the latter could produce higher-fidelity reconstructions.

In this work, we aim to investigate and compare the effects of employing various retrospective subsampling schemes on the quality of DL-based learned image reconstructions. To that end, we trained and tested Recurrent Variational Networks \cite{Yiasemis_2022_CVPR} (RecurrentVarNets) on retrospectively subsampled $k$-space measurements. We performed experiments under either scheme-specific or multi-scheme setups, in which models were trained and evaluated on data subsampled with either individual or multiple, respectively, subsampling schemes.

The contributions and findings of our work can be summarized as follows:
\begin{itemize}[leftmargin=*]
    \item We provide a review of eight currently employed (retrospective) subsampling techniques.  
    \item We experimentally show that DL models trained and evaluated on non-rectilinearly, compared to rectilinearly, subsampled data output superior reconstructions, especially for high acceleration factors.
    \item We demonstrate that models trained on data subsampled with multiple instead of individual patterns, can reconstruct rectilinearly subsampled data with higher fidelity.  
\end{itemize}
\section{Background - Theory}
\label{sec:sec2}

\subsection{MRI Acquisition}
\label{sec:subsec2.1}

MRI reconstruction is an inverse problem on account of the fact that MR scanners acquire MRI measurements in the spatial-frequency domain, also known as the Fourier space, and an inversion procedure is required to produce the desired MR image. 

Let $n = {n_x} \times  {n_y}$ denote the spatial size of the reconstructed data. In the case of single-coil acquisition, the relationship between the underlying (vectorized) image $\vec{x} \, \in \,  \mathbb{C}^{n}$ and the (vectorized) single-channel $k$-space $\vec{y} \, \in \, \mathbb{C}^{n}$ is given by

\begin{equation}
    \vec{y} \, = \, \mathcal{F}(\vec{x}) \, + \, \vec{e},
    \label{eq:singlecoil}
\end{equation}

where $\mathcal{F}$ denotes the two-dimensional (Fast) Fourier Transform (FFT) and $\vec{e} \, \in \, \mathbb{C}^{n}$ some measurement noise.
% , usually assumed additive and normally distributed. 

\subsubsection{Parallel MRI Acquisition}
\label{sec:subsubsec2.1.1}
In PI, multiple receiver coils are placed around the subject to speed up the acquisition. Assuming a number of $n_c$ coils, the acquired (multi-channel) $k$-space measurements are given by
\begin{equation}
\begin{gathered}
        \vec{y} \, = \, \big(\vec{y}^1,\,\cdots,\,\vec{y}^{n_c}\big)\,\in\,\mathbb{C}^{n \times n_c}, \\
        \vec{y}^k \, = \, \mathcal{F}(\mat{S}^k\vec{x}) \, + \, \vec{e}^k, \quad k\,=\,1,\,2,\,\cdots,\,n_{c},
    \label{eq:multicoil}
\end{gathered}
\end{equation}
where $\vec{e}^k$ denotes noise measured by the $k^{\text{th}}$ coil and $\mat{S}^k\,\in\, \mathbb{C}^{n\times n}$  the sensitivity map of the $k^{\text{th}}$ coil expressed as a diagonal complex matrix. Within each coil's reception region, these maps encode their spatial sensitivity by measuring the relative weighting of signals acquired from various locations within the subject. The sensitivity maps are usually normalized as follows
\begin{equation}
    \sum_{k=1}^{n_c} {\mat{S}^k}^{*}{\mat{S}^k}\,=\, \mat{I}_n,
    \label{eq:sense_normalization}
\end{equation}
\noindent
where ${\mat{S}^k}^{*}$ indicates the complex conjugate of ${\mat{S}^k}$ and $\mat{I}_n\,\in\, \mathbb{R}^{n\times n}$ denotes the $n$-rank identity matrix.

Obtaining an image from multi-channel measurements $\vec{y}$ can done by using the root-sum-of-squares (RSS) which operates as follows:
\begin{equation}
\begin{gathered}
    \vec{x}_{\text{rss}}\,=\,\text{RSS}\big(\hat{\vec{x}}^1,\,\cdots,\,\hat{\vec{x}}^{n_c}\big)\,=\,(\sum_{k=1}^{n_c}|\,\hat{\vec{x}}^k\,|^2)^{\frac{1}{2}},\\ 
    \hat{\vec{x}}^k\,=\,\mathcal{F}^{-1}(\vec{y}^k),\quad k\,=\,1,\,2,\,\cdots,\,n_{c}.
    \label{eq:RSS_method}
    \end{gathered}
\end{equation}
% and,
% \begin{equation}
% \begin{gathered}
%     \vec{x}_{\text{sense}}\,=\,\big | \text{SENSE}\big(\hat{\vec{x}}^1,\,\cdots,\,\hat{\vec{x}}^{n_c}\big) \big |\,=\, \big | \sum_{k=1}^{n_c}{\mat{S}^k}^*\hat{\vec{x}}^k \big | \,,\\ 
%     \hat{\vec{x}}^k\,=\,\mathcal{F}^{-1}(\vec{y}^k),\quad k\,=\,1,\,2,\,\cdots,\,n_{c}.
%     \label{eq:SENSE}    
% \end{gathered}
% \end{equation}

%%%%%%%%%%%%%%%%%%%%%%%%%%%%%%%%%%%%%%%%%%%%%%%%%%%%%%%%%%%%%%%%%%%%%

\subsection{Accelerated MRI Acquisition}
\label{sec:subsec2.2}

To accelerate the MRI acquisition, the $k$-space is subsampled by collecting fewer than necessary measurements. The subsampling procedure can be described as the application of a subsampling operator $\mat{U}$ on the fully-sampled $k$-space measurements. The subsampled $k$-space is given by 
\begin{equation}
    \tilde{\vec{y}}^k \, = \, \mat{U}\vec{y}^k \, = \, \mat{U}\mathcal{F}(\mat{S}^k\vec{x}) \, + \,\tilde{\vec{e}}^k,\quad k\,=\,1,\,2,\,\cdots,\,n_{c},
    \label{eq:multicoil_subsampled}
\end{equation}
where $\tilde{\vec{e}}^k = \mat{U}\vec{e}^k$ and $\mat{U}$ is expressed as a binary diagonal mask indicating which measurements are sampled as follows:
\begin{equation}
    \vec{z}_{\mat{U}}\,:=\,(\mat{U}\vec{z})_{i}\,=\,
    \begin{cases} 
      \vec{z}_{i}, & \mat{U}_{ii} \, = \, 1 \\
      0, & \mat{U}_{ii} \, = \, 0.
   \end{cases}
   \label{eq:mask_definition}
\end{equation}
\noindent
The magnitude of the acceleration is determined by an acceleration factor $R$. For a specific $R$, $\mat{U}$ can be chosen such that
\begin{equation}
n\,\cdot\,\big(\sum_{i=1}^{n}\mat{U}_{ii}\big)^{-1} \, \approx \,R.
\end{equation}
\noindent
% Note that the same subsampling is applied to all individual coil data.

%%%%%%%%%%%%%%%%%%%%%%%%%%%%%%%%%%%%%%%%%%%%%%%%%%%%%%%%%%%%%%%%%%%%%

\subsection{Accelerated MRI Reconstruction}
\label{sec:subsec2.3.1}
\subsubsection{Sensitivity Map Estimation}
\label{sec:subsubsec2.3.1}

The sensitivity maps $\mat{S} \, = \, (\mat{S}^1,\, \cdots,\, \mat{S}^{n_c})$ can be estimated by various methods found in the literature \cite{sense, Griswold_2002, isense, Uecker2014-vm}. A common method for estimating them is by fully-sampling a small region of the center of the $k$-space, also known as the autocalibration signal (ACS) which includes low frequencies \cite{Yiasemis_2022_CVPR, varnet}. 

Let $\mat{U}_{\text{ACS}}$ denote the ACS-subsampling operator such that when applied on $k$-space data it outputs the fully-sampled ACS region, i.e.:
\begin{equation}
     \vec{z}_{\text{acs}}\,:=\,(\mat{U}_\text{ACS}\vec{z})_{i}\,=\,
    \begin{cases} 
      \vec{z}_{i}, & i \in \text{ ACS region} \\
      0, & \text{otherwise}.
   \end{cases}
   \label{eq:acs_definition}
\end{equation}

Subsequently, to obtain an initial approximation of the sensitivity maps we use the RSS method:
\begin{equation}
    \tilde{\mat{S}}^k\,\approx\,\diag\Big[\vec{x}_{\text{acs}}^{k} \, \oslash \, \vec{x}_{\text{acs}} \Big],\quad k\,=\,1,\,2,\,\cdots,\,n_{c},
    \label{eq:sens_map_estimation}
\end{equation}
where $\oslash$ denotes the element-wise division, and
\begin{equation}
    \vec{x}_{\text{acs}}^{k}\,= \, \mathcal{F}^{-1} \big( \mat{U}_\text{ACS}\tilde{\vec{y}}^k \big), \quad \vec{x}_{\text{acs}}\,=\,\text{RSS}\Big(\Big \{\vec{x}_{\text{acs}}^{k}\Big\}_{k=1}^{n_c} \Big).
\end{equation}
\noindent
% For our experiments we used \Eq{sens_map_estimation} to initially estimate the sensitivity maps for each case, but as we will discuss later we use a DL-based method to better refine them.
%%%%%%%%%%%%%%%%%%%%%%%%%%%%%%%%%%%%%%%%%%%%%%%%%%%%%%%%%%%%%%%%%%%%%

\subsubsection{Accelerated MRI Reconstruction as an Inverse Problem}
\label{sec:subsubsec2.3.2}
Obtaining a reconstruction from accelerated multicoil $k$-space measurements is an inverse problem, with a forward problem given by \Eq{multicoil_subsampled}. We can rewrite \Eq{multicoil_subsampled} in a more compact notation:
\begin{equation}
    \vec{\tilde{y}}\,=\, \mathcal{A}_{\mat{U},\mat{S}}(\vec{x}),\quad \mathcal{A}_{\mat{U},\mat{S}}\,:=\,\mat{U}\,\circ\,\mathcal{F}\,\circ\,\mathcal{E}_{\mat{S}},
    \label{eq:multicoil_op}
\end{equation}
where $\mathcal{A}_{\mat{U},\mat{S}}:\mathbb{C}^{n}\rightarrow\mathbb{C}^{n\times n_c}$ denotes the forward operator and $\mathcal{E}_{\mat{S}}:\mathbb{C}^{n}\rightarrow\mathbb{C}^{n\times n_c}$ is called the expand operator which maps an image $\vec{x}\in\mathbb{C}^{n}$ to the individual coil images using $\mat{S}$:
\begin{equation}
    \mathcal{E}_{\mat{S}}(\vec{x})\,=\,\big(\mat{S}^1\vec{x},\,\cdots,\,\mat{S}^{n_c}\vec{x}\big).
    \label{eq:expand_op}
\end{equation}

\noindent
The backward operator of $\mathcal{A}_{\mat{U},\mat{S}}$ is given by 
\begin{equation}
    \mathcal{A}_{\mat{U},\mat{S}}^{*}\,:=\,\mathcal{R}_{\mat{S}}\,\circ\,\mathcal{F}^{-1}\,\circ\,\mat{U}:\,\mathbb{C}^{n\times n_c}\rightarrow\mathbb{C}^{n},
    \label{eq:multicoil_back_op}
\end{equation}
where  $\mathcal{R}_{\mat{S}}:\mathbb{C}^{n\times n_c}\rightarrow\mathbb{C}^{n}$ is called the reduce operator that combines individual coil images $\vec{z}\,\in\,\mathbb{C}^{n\times n_c}$ using $\mat{S}$ as follows:
\begin{equation}
    \mathcal{R}_{\mat{S}}(\vec{z})\,=\,\sum_{k=1}^{n_c}{\mat{S}^k}^{*}\vec{z}^{k}.
    \label{eq:reduce_op}
\end{equation}
\noindent
Note that the operators $\mat{U}$, $\mathcal{F}$ and $\mathcal{F}^{-1}$ in \Eq{multicoil_op} and \Eq{multicoil_back_op} are applied coil-wise.

Subsampling the $k$-space causes obtaining a solution to \Eq{multicoil_op} to be an ill-posed inverse problem \cite{5163622, 10.48550}, and therefore, a solution through direct inversion is not feasible. In CS, recovering an estimation of the ground truth image $\vec{x}$ from the subsampled MRI measurements $\tilde{\vec{y}}$ can be formulated as a solution to a variational optimization problem as follows:
\begin{equation}
	\hat{\vec{x}}\,=\,\argmin_{\vec{w}}\, \big | \big | \mathcal{A}_{\mat{U},\mat{S}}(\vec{w})\,-\,\tilde{\vec{y}}\big |\big |_2^2 \,+\,\alpha\, \mathcal{G}(\vec{w}),
	\label{eq:var_problem}
\end{equation}
where $\mathcal{G}\,:\,\mathbb{C^{n}}\rightarrow\mathbb{R}$ is a regularization function which can impose prior information about the solution and $\alpha\,>\,0$ is a regularization parameter.  In the literature various choices of $\mathcal{G}$ and algorithms for solving \Eq{var_problem} have been employed \cite{Sodickson_2001, PICCOLOMINI2002325, Lin_2004, https://doi.org/10.1002/mrm.22964}. 

%%%%%%%%%%%%%%%%%%%%%%%%%%%%%%%%%%%%%%%%%%%%%%%%%%%%%%%%%%%%%%%%%%%%%

\subsubsection{Deep Learning-based Accelerated MRI Reconstruction}
\label{sec:subsubsec2.3.3}

With the advent of the involvement of DL in MRI reconstruction tasks, the need for handcrafting a specific regularization function has been replaced with CNN-based architectures. A plethora of approaches solve \Eq{var_problem} by unrolling it into a gradient descent iterative optimization scheme over $T$ time-steps:
\begin{equation}
\begin{gathered}
    \vec{w}_{t+1}\,=\,\vec{w}_{t}\,-\,\alpha_{t+1}\,\mathcal{A}_{\mat{U},\mat{S}}^{*}\big(\mathcal{A}_{\mat{U},\mat{S}}(\vec{w}_{t})\,-\,\tilde{\vec{y}}\big)\,+\, \mathcal{H}_{\theta_{t+1}}(\vec{w}_{t}),\\
    t\,=\,0,\,\cdots,\,T-1,
    \label{eq:unrolled_var_problem}
\end{gathered}
\end{equation}
where $\alpha_{t}$ denotes a (trainable) learning rate and $\mathcal{H}_{\theta_t}$ a CNN-based architecture with trainable parameters $\theta_t$. The initial image $\vec{w}_0$ in \Eq{unrolled_var_problem} can be chosen as a zero-filled reconstruction using $\tilde{\vec{y}}$. 

Sensitivity maps can be estimated as in \Eq{sens_map_estimation} and/or can be refined using another CNN-based model $\mathcal{S}_{\psi}$ with trainable parameters $\psi$ which takes as input the estimation $\tilde{\mat{S}}\,=\,(\tilde{\mat{S}}^1, \,\cdots,\,\tilde{\mat{S}}^{n_c})$ as in \Eq{sens_map_estimation}:
\begin{equation}
    \mat{S}\,=\,\mathcal{S}_{\psi}(\tilde{\mat{S}}).
\end{equation}

Optimization of \Eq{unrolled_var_problem} may alternatively be performed in the $k$-space domain as demonstrated by some authors \cite{varnet, Yiasemis_2022_CVPR}:
\begin{equation}
    \begin{aligned}
    \vec{y}_{t+1} &= \vec{y}_{t} - \alpha_{t+1} \, \mat{U}\big(\vec{y}_{t} - \tilde{\vec{y}}\big) \\
    &\quad + \mathcal{F} \circ \mathcal{E}_{\mat{S}} \circ \mathcal{H}_{\theta_{t+1}} \circ \mathcal{R}_{\mat{S}} \circ \mathcal{F}^{-1}\big(\vec{y}_{t}\big),
    \end{aligned}
    \label{eq:unrolled_var_problem_kspace}
\end{equation}
with initial condition $\vec{y}_0 = \tilde{\vec{y}}$.
%%%%%%%%%%%%%%%%%%%%%%%%%%%%%%%%%%%%%%%%%%%%%%%%%%%%%%%%%%%%%%%%%%%%%

\subsection{\textit{k}-space Sampling}
\label{sec:subsubsec2.4}
\begin{figure*}[!tbh]
\centering
\includegraphics[width=\textwidth]{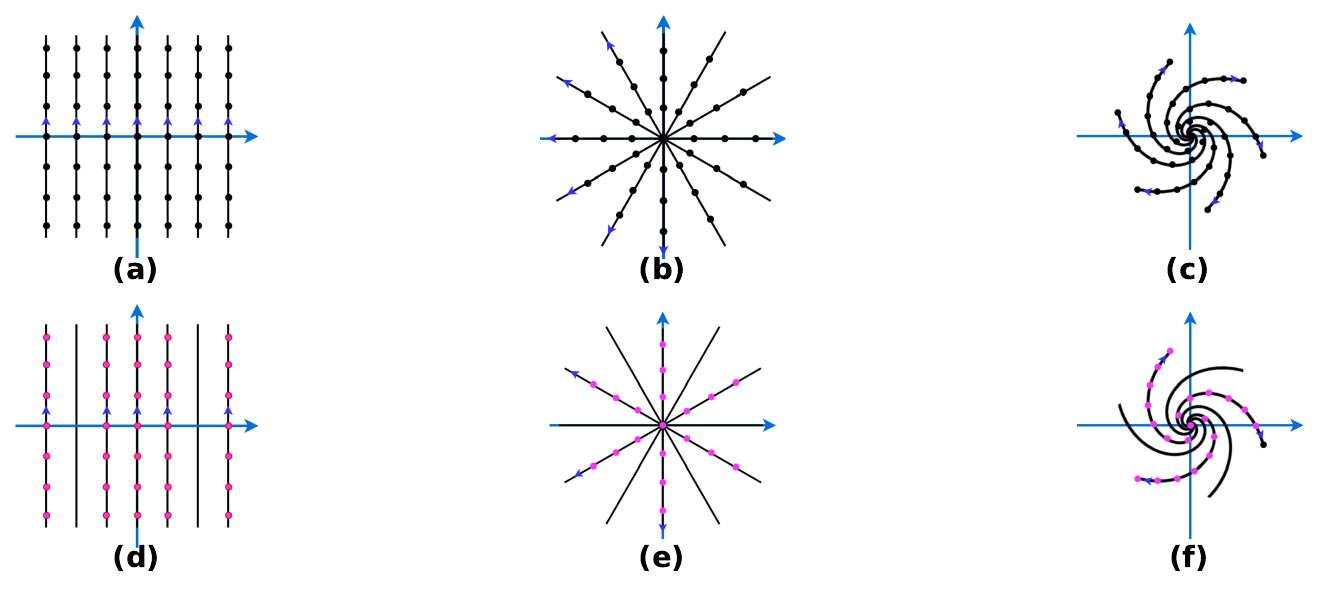}
\caption{\textbf{Top:} $k$-space sampling trajectories. Cartesian: (a) Rectilinear: $k$-space is filled in a line-by-line scheme. Non-Cartesian: (b) Radial: $k$-space is filled with radial spokes passing through the center, (c) Spiral: $k$-space is filled by one or multiple hellical curves.  Each line in (a) and (b), and curve in (c), represent a separate filling. The dark blue arrows show the direction of each readout. \textbf{Bottom:} Subsampled $k$-space trajectories for different acceleration factors: (d) Rectilinear, $R\,\approx\,1.5$. (e) Radial, $R\,\approx\,2$. (f) Spiral, $R\,\approx\,2$.}
\label{fig:trajectories}
\end{figure*}

A sampling scheme or $k$-space trajectory refers to the course of filling up a complex array with $k$-space frequencies acquired over a sequence of time-steps during the MRI acquisition. There exist a wide range of 2D $k$-space trajectories implemented in clinical settings which can be split into two groups: Cartesian and non-Cartesian trajectories. 

Cartesian trajectories aim in collecting samples on a Cartesian or equispaced and rectangular grid. The most common Cartesian trajectory is the rectilinear one in which $k$-space samples are acquired in a line-by-line scheme as illustrated by Figure \hyperref[fig:trajectories]{1(a)} with resulting samples being equidistant in both axes. Note that usually in the literature the Cartesian rectilinear trajectory is referred to as simply \textit{Cartesian}. In this work we use the characterization \textit{Cartesian} to refer to any trajectory acquired on a Cartesian grid.

Non-Cartesian trajectories include schemes such as the radial or the spiral. In the former, $k$-space signal samples are acquired along several spokes crossing its center (see Figure \hyperref[fig:trajectories]{1(b)}), resulting in the center being sampled multiple times. Meanwhile, the latter includes acquiring data on helical curves starting from the center of the $k$-space (see Figure \hyperref[fig:trajectories]{1(c)}). In non-Cartesian trajectories, $k$-space measurements are acquired on a non-Cartesian grid and are therefore not equidistant with each other. For instance, in the radial filling, samples closer to the center are more dense compared to samples further on the radial spokes.

To accelerate the MRI acquisition, the k-space is subsampled by an acceleration factor $R$, resulting in fewer measurements being collected than those strictly required by the Nyquist-Shannon criterion for perfect reconstruction. This can lead to a degradation of the quality of the reconstruction, depending on the magnitude of the acceleration factor. For instance, for $R\,=\,2$, half of the necessary $k$-space measurements are acquired. In Figure \hyperref[fig:trajectories]{1(d)-(f)} we provide examples of subsampled $k$-space trajectories: Cartesian rectilinear (Figure \hyperref[fig:trajectories]{1(d)}), radial and spiral (Figure \hyperref[fig:trajectories]{1(e)-(f)}).

\section{Methods}
\label{sec:sec3}

\subsection{Retrospective \textit{k}-space Subsampling}
\label{sec:subsec3.1}

\begin{figure*}[!tbh]
\centering
\includegraphics[width=1\textwidth]{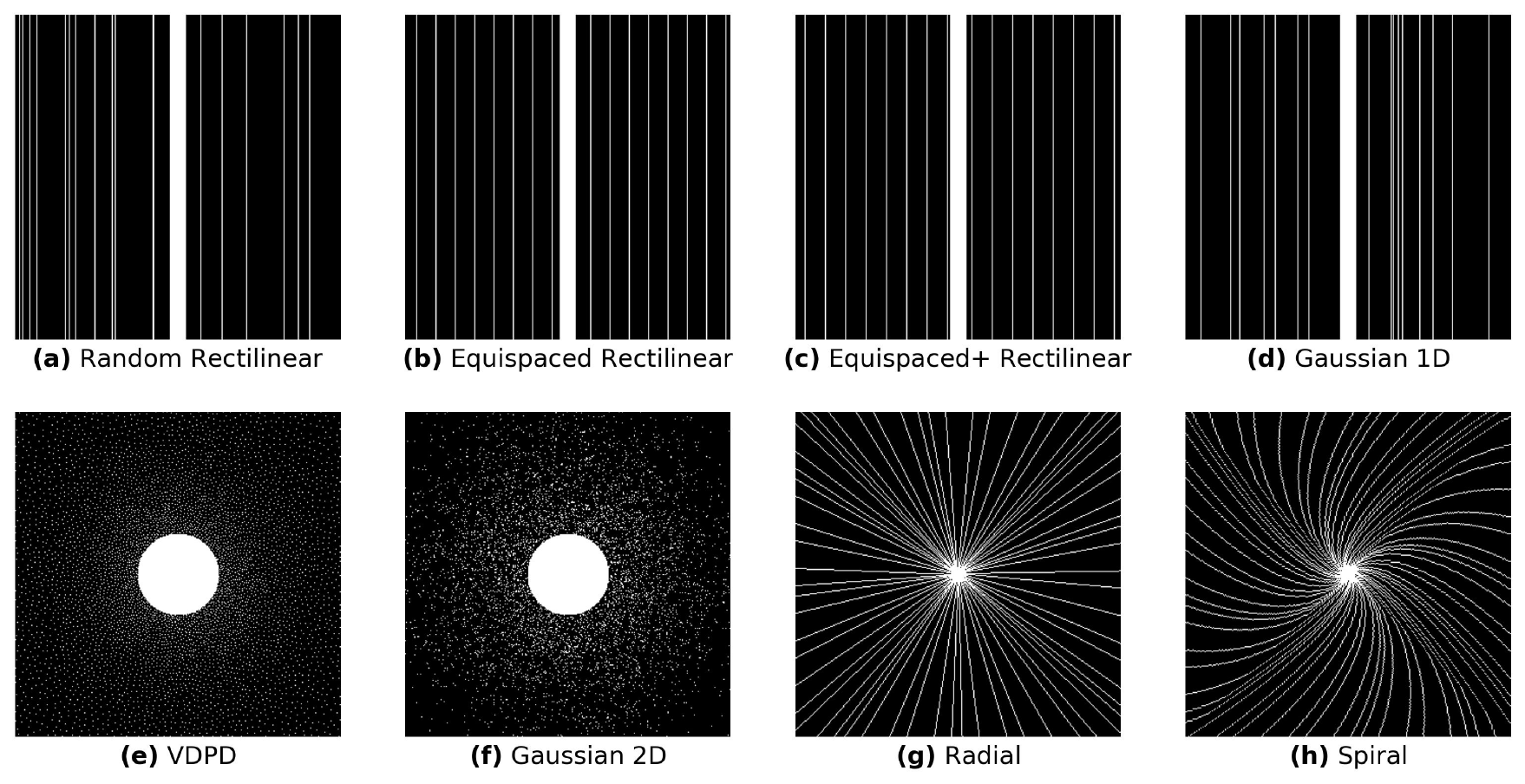}
\caption{Examples of subsampling masks for Cartesian data for acceleration factor $R\,=\,5$. \textbf{(a)-(d)} Rectilinear: generated by first selecting a fraction of center (ACS) columns. The rest of the columns are: (a) selected uniformly at random, (b) equispaced with a fixed distance, (c)  equispaced but symmetric, (d)  selected from the Gaussian distribution \textbf{(e)} Variable-density Poisson Disk: generated using Bridson's fast algorithm \cite{poisson}. \textbf{(f)} Gaussian 2D: samples selected from a 2D Gaussian distribution. For (e)-(f) the ACS region was generated by fully-sampling a centered disk. \textbf{(g)-(h)} Simulated non-Cartesian patterns using the CIRCUS algorithm \cite{circus}: (g) Radial, (h) Spiral.}
\label{fig:masks}
\end{figure*}

The data used in this work consisted of volumes of fully-sampled raw $k$-space measurements acquired on a Cartesian grid. To simulate prospective subsampling we generated 2D subsampling masks which we retrospectively applied onto the fully-sampled multi-coil $k$-space data to produce subsampled/masked measurements. The generated masks were binary, signifying that a sample from the fully-sampled data was masked if and only if the corresponding mask entry was zero.

As in this work we are interested in studying the role of the subsampling pattern on the quality of DL-based reconstructions of subsampled MRI data, we focused on the following retrospective subsampling patterns on the Cartesian grid:

\paragraph{Cartesian Subsampling}
\label{sec:para3.1.1}
\hspace{0pt} \\
Let $0<r_\text{acs}<1$ be the autocalibration region ratio.

\begin{itemize}[leftmargin=*]
    \item Rectilinear: Achieved by including some and omitting other horizontal (phase encoding direction) lines on the Cartesian grid. For the ACS region we used a number of $r_\text{acs} \cdot n_y$ lines. 
    
    We employed four distinct rectilinear sampling patterns: 
    \begin{itemize}[leftmargin=*]
        \item Random (Figure \hyperref[fig:masks]{2(a)}): Lines were drawn from the uniform distribution with possible overlap with the ACS lines.
        \item Equispaced (Figure \hyperref[fig:masks]{2(b)}): Lines were included with a fixed distancing that satisfied the desired acceleration.
        \item Equispaced+ (Figure \hyperref[fig:masks]{2(c)}): Modified equispaced pattern that exploits $k$-space symmetry as presented in \citenum{defazio_2019}.
        \item Gaussian 1D (Figure \hyperref[fig:masks]{2(d)}): Lines were drawn from the Gaussian distribution with mean $\mu=\frac{n_y}{2}$ and standard deviation $\sigma=4 \cdot \sqrt{\mu}$.
    \end{itemize}
    \item Variable Density Poisson Disk (VDPD, Figure \hyperref[fig:masks]{2(e)}): Combines both random, and denser center sampling. For our implementation we followed Bridson's algorithm \cite{poisson}. Samples were drawn with a density $\frac{1}{1\,+\,s\cdot|r|}$, i.e., inversely proportional to the $k$-space radius $r$ and a slope $s$ which were determined by the prescribed acceleration.
    
    \item Gaussian 2D (Figure \hyperref[fig:masks]{2(f)}): Samples were drawn from the Gaussian distribution with mean $\boldsymbol{\mu}=\frac{1}{2}(n_x,n_y)$ and covariance $\mat{\Sigma}=4\cdot\mat{I}_2\cdot\sqrt{\boldsymbol{\mu}}^T$. 

\end{itemize}
Code for VDPD and Gaussian 2D schemes was implemented in Cython for fast and efficient sampling, and for the autocalibration signal we fully-sampled a centered disk with a radius $\frac{n \cdot r_{\text{acs}}}{\pi}$. 

\paragraph{Simulated non-Cartesian Subsampling}
\label{sec:para3.1.2}
 \hspace{0pt} \\
To simulate non-Cartesian subsampling, we applied the CIRcular Cartesian UnderSampling (CIRCUS) \cite{circus} technique to produce the following retrospective subsampling patterns for Cartesian data:
\begin{itemize}[leftmargin=*]
    \item Radial (Figure \hyperref[fig:masks]{2(g)}): Simulates radial subsampling on the Cartesian grid.
    \item Spiral (Figure \hyperref[fig:masks]{2(h)}): Simulates spiral subsampling on the Cartesian grid.
\end{itemize}

For randomization CIRCUS' offset parameter as defined in \citenum{circus}, can be set to produce random radial and spiral patterns. CIRCUS was modified to output masks by specifying the acceleration factor. Note that in contrast to the rest of the subsampling patterns above, for the non-Cartesian case we did not sample the ACS region exclusively, as these patterns already fully-sample a great portion of the $k$-space center. Therefore, for the ACS subsampling mask $\mat{U}_{\text{acs}}$ we calculated the largest sampled centered disk from $\mat{U}$.

%%%%%%%%%%%%%%%%%%%%%%%%%%%%%%%%%%%%%%%%%%%%%%%%%%%%%%%%%%%%%%%%%%%%%%%%%%%%%%%%%%%%%%%%%%%%%%%%%%%%%%%%%%%%%%%%%
\subsection{Deep MRI-reconstruction Model Architecture}
\label{sec:subsec3.2}

\subsubsection{The Recurrent Variational Network}
\label{sec:subsec3.2.1}

To compare and evaluate the aforementioned subsampling techniques we employed a DL-based reconstruction network, namely the Recurrent Variational Network \cite{Yiasemis_2022_CVPR} (RecurrentVarNet). The RecurrentVarNet is a DL-based inverse problem solver previously applied on the task of Accelerated MRI Reconstruction \cite{Yiasemis_2022_CVPR} with state-of-the-art performance (MC-MRI reconstruction challenge winning solution \cite{Beauferris_2022}). It iteratively solves the gradient descent scheme in the measurements domain as portrayed by \Eq{unrolled_var_problem_kspace} using convolutional recurrent neural networks (ConvRNNs) as a regularizer. The RecurrentVarNet takes subsampled multi-coil $k$-space measurements as input and outputs a prediction of the fully-sampled multi-coil $k$-space. It comprises three main modules:
\begin{itemize}[leftmargin=*]
    \item \textbf{Recurrent Variational Block (RecurrentVarNet Block).} The RecurrentVarNet consists of $T$ RecurrentVarNet Blocks, which are the main blocks of the method, each responsible for performing an unrolled gradient descent optimization time-step as in \Eq{unrolled_var_problem_kspace} by replacing $\mathcal{H}_{\theta_{t}}$ with a recurrent unit, denoted as $\text{RNN}_{\theta_{t}}$:
    \begin{equation}
        \begin{gathered}
        \vec{w}_{t},\,\vec{h}_{t+1}\,=\,\text{RNN}_{\theta_{t+1}}\bigg(\mathcal{R}_{\mat{S}} \, \circ \, \mathcal{F}^{-1}\big(\vec{y}_{t}\big),\,\vec{h}_{t}\bigg),\\
         \vec{y}_{t+1}\,=\,\vec{y}_{t}\,-\,\alpha_{t+1} \, \mat{U}\big(\vec{y}_{t}\,-\,\tilde{\vec{y}}\big)\,+\, \mathcal{F} \, \circ \, \mathcal{E}_{\mat{S}} \big(\vec{w}_{t}\big),\\
         \vec{y}_0\,=\,\tilde{\vec{y}}, \quad t\,=\,0,\,\cdots,\,T-1.
        \end{gathered}
    \label{eq:recurrentvarnet_block}
    \end{equation}
    Each $\text{RNN}_{\theta_{t}}$  is consisted of a convolutional layer (Conv) with a $5\times 5$ kernel followed by $n_l$ cascades of alternating Convs with a $3\times 3$ kernel and convolutional gated recurrent units (ConvGRUs). A rectified linear unit is applied after each Conv excluding the last. $\text{RNN}_{\theta_{t}}$ takes as input intermediate quantities of the image projection of the refined $k$-space $\mathcal{R}_{\mat{S}} \, \circ \, \mathcal{F}^{-1}\big(\vec{y}_{t-1}\big)$ and the hidden state $\vec{h}_{t-1}$ from the previous time-step.
    
    \item \textbf{Recurrent State Initializer (RSI).} It produces an initialization for the first hidden state $\vec{h}_0$ to be used by $\text{RNN}_{\theta_{1}}$ provided as input the SENSE reconstruction of the image projection of $\vec{y}_0$:
    \begin{equation}
        \vec{h}_0\,=\,\text{RSI}\bigg(\mathcal{R}_{\mat{S}} \circ \mathcal{F}^{-1}(\vec{y_0}) \bigg).
    \end{equation}
    \item \textbf{Sensitivity Estimation - Refinement (SER).} It estimates at each iteration the coil sensitivity maps as in \Eq{sens_map_estimation}, and refines them using a U-Net \cite{Unet} with trainable parameters $\psi$ denoted as $\mathcal{S}_{\psi}$:
    \begin{equation}
         \mat{S}\,=\,\text{SER}\big(\tilde{\mat{S}}\big):  \quad \mat{S}^{k}\,=\,\mathcal{S}_{\psi}\big(\tilde{\mat{S}^{k}}\big), \quad k\,=\,1,\,\cdots,\,n_c.
    \label{eq:recurrentvarnet_rsi}
\end{equation}
\end{itemize}

\subsection{Experimental Setup}
\label{sec:subsec3.3}

\begin{figure*}[!ht]
\centering
\includegraphics[width=0.96\textwidth]{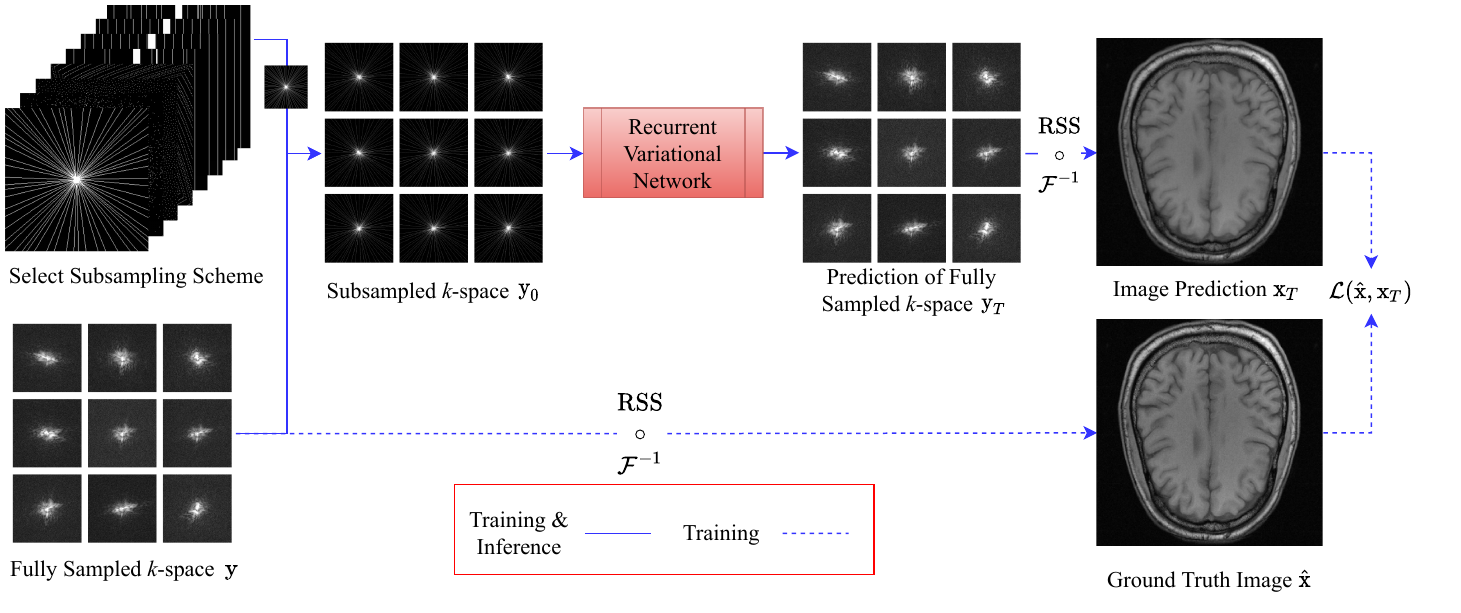}
\caption{Experiments pipeline: For each subsampling scheme (e.g. here: radial), the fully-sampled multi-coil $k$-space is retrospectively subsampled and used as input to a RecurrentVarNet, which outputs a prediction of the fully-sampled $k$-space. The predicted $\Vec{x}_T$ and ground truth  $\hat{\Vec{x}}$ images are produced by applying the $\text{RSS} \circ \mathcal{F}^{-1}$ operator onto $\Vec{y}_T$ and $\Vec{y}$, respectively. During training, the loss $\mathcal{L}$ is calculated using $\Vec{x}_T$ and $\hat{\Vec{x}}$.}    
\label{fig:pipeline}
\end{figure*}
To perform our experiments, we retrospectively subsampled the fully-sampled $k$-space data by generating subsampling masks as introduced in \Section{subsec3.1}. We carried out two classes of experiments: scheme-specific and multi-scheme. An overview of our experimental setup is illustrated in \Figure{pipeline}.

\subsubsection{Scheme-specific Setup}
\label{sec:subsubsec3.3.1}
To compare the individual subsampling patterns and demonstrate their effect on the quality of DL-based reconstruction, we first performed experiments in a scheme-specific setting:  for each dataset-pattern pair we ran individual experiments by training and evaluating (twenty-four) distinct RecurrentVarNets (with the same choice of hyper-parameters as outlined in \Section{para3.3.4.1}).

\subsubsection{Multi-scheme Setup}
\label{sec:subsubsec3.3.2}
In the multi-scheme setting, our goal was twofold: Firstly, we aimed to investigate further the effect of each subsampling scheme on the quality of DL-based reconstruction. Secondly, and most importantly, we wanted to asses whether or not a DL-based model trained in a multi-scheme fashion (i.e. training measurements subsampled with multiple subsampling patterns) demonstrated higher reconstruction performance compared to being trained in a scheme-specific fashion (as in \Section{subsubsec3.3.1}). Therefore, for each dataset, a RecurrentVarNet was trained on data arbitrarily subsampled with any of the presented subsampling schemes in \Section{subsec3.1} and evaluated on all of them. 

\subsubsection{Subsampling}
\label{sec:subsubsec3.3.3}
In both, scheme-specific and multi-scheme settings, throughout the training phase, subsampling masks were generated with an acceleration factor of $R\,=\,$2, 4 or 8, and were retrospectively applied onto the fully-sampled data. At validation and testing times, data were 2-fold, 4-fold, and 8-fold retrospectively subsampled. 

For the Cartesian masks we set $r_\text{acs} = 0.16, 0.08, 0.04$  for $R\,=\,2, 4, 8$, respectively.

Note that for all experiments, subsampling masks during training were randomly generated to maximize the amount of data that each model saw, while for validation and testing they were predetermined. Additionally, the same subsampling mask was applied to all slices of each volume during validation and testing. It should also be highlighted that all individual coil data of each sample were subsampled with the same subsampling mask as this reflected clinical subsampling.

\subsubsection{Model Implementation Details}
\label{sec:subsubsec3.3.4}

\paragraph{Hyper-parameter Choice}
\label{sec:para3.3.4.1}
\hspace{0pt} \\
For the RecurrentVarNets we used $T=8$ time-steps. For each RecurrentVarNet Block we used $n_l=4$ alternating cascades and for the number of filters in each Conv and hidden size in each ConvGRU we chose $n_f=256$ channels. For the implementation of the RSI and SER modules we picked the same choice of hyper-parameters as the original paper \cite{Yiasemis_2022_CVPR}.

\paragraph{Training \& Optimization Details}
\label{sec:para3.3.4.2}
\hspace{0pt} \\
For training and optimization we utilized PyTorch \cite{pytorch}. All components of the Recurrent Variational Network were trained end-to-end and optimization was performed utilizing the Adam algorithm with coefficients $(\beta_1,\,\beta_2)=(0.9,\,0.999)$, stability parameter $\epsilon=1 \times 10^{-8}$ and no weight decay. 

Experiments were performed utilizing NVIDIA RTX A6000, Quadro RTX 8000, or A100 GPUs. Models were trained to convergence with a batch size of 1 slice multi-coil $k$-space data. The total number of trainable parameters for each model amounted to approximately 27,626k.

\paragraph{Training Loss Function}
\label{sec:para3.3.4.3}
\hspace{0pt} \\
At each training iteration each model was fed with subsampled multi-coil $k$-space measurements $\tilde{\vec{y}}$ and produced a prediction $\vec{y}_T$ of the fully-sampled reference $k$-space $\vec{y}$. Loss was computed in the image domain using  $\vec{x}_T = \text{RSS}\circ\mathcal{F}^{-1}(\vec{y}_T)$ as the image prediction and $\hat{\vec{x}} = \text{RSS}\circ\mathcal{F}^{-1}(\vec{y})$ as the image reference.

As a loss function we used a combination of the mean average error (MAE) loss $\mathcal{L}_\text{MAE}$ and the structural similarity index measure (SSIM) loss $\mathcal{L}_\text{SSIM}$:
\begin{equation}
\begin{split}
    \mathcal{L}(\hat{\vec{x}}, \vec{x}_T)\, &= \,\mathcal{L}_\text{MAE}(\hat{\vec{x}}, \vec{x}_T)\,+\, \mathcal{L}_\text{SSIM}(\hat{\vec{x}}, \vec{x}_T) \, \\
    &= \, \big|\big|\hat{\vec{x}}\,-\, \vec{x}_T\big|\big|_1\,+\,\big(1\,-\,\text{SSIM}(\hat{\vec{x}}, \vec{x}_T)\big),
\end{split}
\end{equation}
where SSIM is defined in \Section{para3.3.6.1}.

\subsubsection{Datasets}
\label{sec:subsubsec3.3.5}

To perform our experiments we employed three open source datasets, the fastMRI knee and brain datasets \cite{fastmridataset} which are to-date the largest publicly available MRI datasets, and the Calgary-Campinas (CC) brain dataset which was released as part of the Multi-Coil MRI Reconstruction Challenge \cite{Beauferris_2022}. All datasets consisted of raw $k$-space volumes of 2D multi-coil, fully-sampled Cartesian $k$-space data. The acquisition parameters and the splitting ratios we opted for in our experiments are summarized in \Table{A1}.

\subsubsection{Quality Analysis}
\label{sec:subsec3.3.6}

\paragraph{Metrics}
\label{sec:para3.3.6.1}
To analyse and compare our results we employed five evaluation metrics commonly used in image processing. Let $\vec{u}\,\in\,\mathbb{R}^{n}$ be the reference image and $\vec{v}\,\in\,\mathbb{R}^{n}$ the prediction, they are defined as follows:

\begin{description}[style=unboxed,leftmargin=0cm]
\item[A)]   \textbf{Structural Similarity Index Measure (SSIM)}
\begin{equation}
    \text{SSIM}(\vec{u},\,\vec{v}) =
    \frac{1}{M}\sum_{i=1}^{M} \frac{(2\mu_{\vec{u}_i}\mu_{\vec{v}_i} + c_1)(2\sigma_{\vec{u}_i\vec{v}_i} + c_2)}{({\mu^2_{\vec{u}_i}} +{\mu^2_{\vec{v}_i}} + c_1)({\sigma^2_{\vec{u}_i}} + {\sigma^2_{\vec{v}_i}} + c_2)},
    \label{eq:ssim}
\end{equation}
    
where $\vec{u}_i,\,\vec{v}_i,\,i=1,\cdots,M$ are image windows of size $(w_x, w_y)= (7, 7)$ from $\vec{u}$ and $\vec{v}$, respectively. The numbers $\mu_{\vec{u}_i}$, $\mu_{\vec{v}_i}$, $\sigma_{\vec{u}_i}$ and $\sigma_{\vec{v}_i}$ denote the means and standard deviations of each image window and $\sigma_{\vec{u}_i\vec{v}_i}$ denotes the covariance between $\vec{u}_i$ and $\vec{v}_i$. The constants $c_1 = 0.01$ and $c_2 = 0.03$ are used for numerical stability. 

\item[B)] \textbf{Peak Signal-to-Noise Ratio (pSNR)}
\begin{equation}
    \text{pSNR}(\vec{u},\, \vec{v})\, = \, 20\log_{10}\bigg(\frac{\max(\vec{u})}{{\,\sqrt{\frac{1}{n}\sum_{i}^n(\vec{u}_{i} - {\vec{v}}_{i})^2}}}\bigg).
    \label{eq:psnr}
\end{equation}

\item[C)] \textbf{Signal-to-Noise Ratio (SNR)}
\begin{equation}
    \text{SNR}(\vec{u},\, \vec{v})\, = \, 10\log_{10}\bigg(\frac{\sum_{i}^n\vec{u}_{i}^2}{{\,\sum_{i}^n(\vec{u}_{i} - {\vec{v}}_{i})^2}}\bigg).
    \label{eq:snr}
\end{equation}

\item[D)] \textbf{High Frequency Error Norm (HFEN)}
\begin{equation}
    \text{{HFEN}}_C(\vec{u},\, \vec{v})\, = \, \frac{|| \text{{LoG}}(\vec{u}) - \text{{LoG}}(\vec{v}) ||_C}{||\text{{LoG}}(\vec{v})||_C},
    \label{eq:hfen}
\end{equation}
where where $\text{LoG}$ is a Laplacian-of-Gaussian filter  \cite{5617283} with kernel of size $15\times 15$ and with a standard deviation of 2.5, and $C\,=\, 1 \text{or } 2$.

\item[E)] \textbf{Normalized Mean Squared Error (NMSE)}
\begin{equation}
    \text{NMSE} (\vec{u},\, \vec{v})\,= \, \frac{||\vec{u}\,-\,\vec{v}||_2^2}{||\vec{u}||_2^2}\,= \, \frac{\sum_{i}^n(\vec{u}_{i} - {\vec{v}}_{i})^2}{\sum_{i}^n\vec{u}_{i}^{2}}.
    \label{eq:nmse}
\end{equation}
\end{description}
\noindent
Note that the higher the computed SSIM, SNR, and pSNR values, the higher the quality of the reconstruction is. Conversely, for the HFEN and NMSE values, the lower they are, the higher the quality of the reconstruction is. Reported values for SSIM and NMSE are multiplied by 100 and 1000, respectively.

\subsubsection{Significance Tests}
\label{sec:subsubsec3.3.7}   
To perform significance tests we used the Almost Stochastic Order (ASO) test \cite{dror-etal-2019-deep, ulmer2022deep} with a 95\% confidence level ($\alpha=0.05$). Each ASO test outputs a violation error $\epsilon_{\text{min}}$ which denotes the degree to which the hypothesis that "method A is always better than method B" is being violated. If $\epsilon_{\text{min}} < 0.5$ one can claim that A is better than B, and otherwise if $\epsilon_{\text{min}}\ge0.5$.

\subsubsection{Robustness to Model Choice Experiments}
\label{sec:subsubsec.3.3.8}
To ensure that the findings presented in Section \Section{sec4} are not reliant on the choice of the model architecture, we conducted additional experiments using an alternative advanced deep-MRI reconstruction network known as the Recurrent Inference Machine (RIM). Specifically, we replicated the scheme-specific experiments on the CC dataset by substituting the RecurrentVarNet with a RIM. The hyper-parameter settings for each RIM were consistent with those employed in the study in \cite{Yiasemis_2022_radial}.
% In order to demonstrate that the findings presented in Section \Section{sec4} are not dependent on the selection of the model architecture, we conducted additional experiments using another advanced deep-MRI reconstruction network called the Recurrent Inference Machine (RIM). Specifically, we replicated the scheme-specific experiments on the CC dataset by substituting the RecurrentVarNet with a RIM. The hyper-parameter settings for each RIM were the same as those used in the study in \cite{Yiasemis_2022_radial}.
\section{Results}
% \newgeometry{top=5mm, bottom=5mm}  

\begin{figure*}[!ht]
    \centering
    \includegraphics[width=0.922\textwidth]{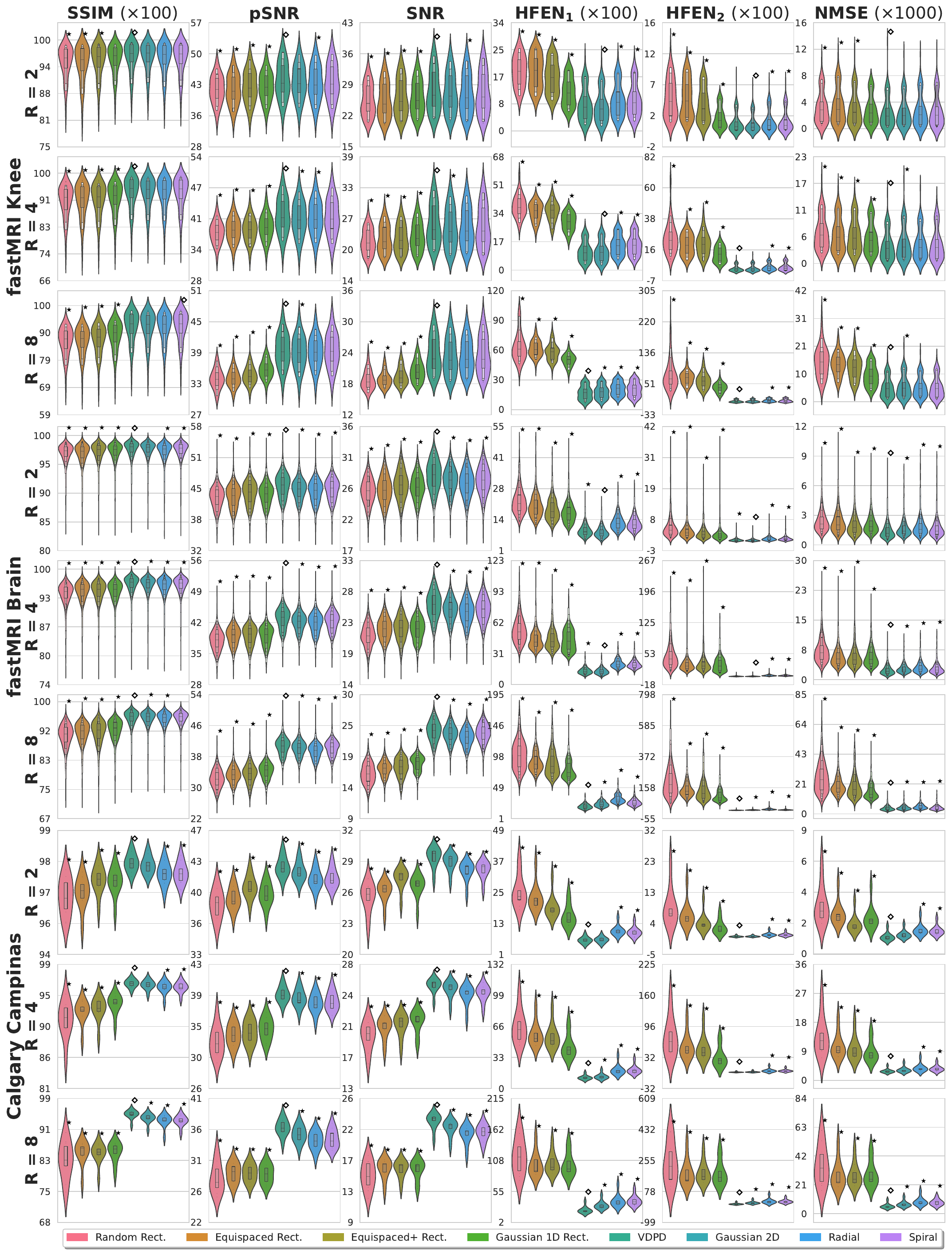}
    \caption{Scheme-specific experiments quantitative results on the test sets. For each dataset-subsampling scheme pair a distinct model was trained (in total 24 models). 
    For each dataset-acceleration-metric combination, pair-wise ASO significance tests were performed between the average best performing, denoted by $\diamond$, and the rest schemes. The symbol $\star$ indicates that VDPD was found to be significantly better. Average metrics are reported in \Table{A2}.
    }
    \label{fig:metrics}
\end{figure*}
% \restoregeometry   

\label{sec:sec4}
To obtain our results, data preparation, retrospective subsampling generation, and model training we used the Deep Image Reconstruction Toolkit (DIRECT) \cite{DIRECTTOOLKIT}.

\subsection{Scheme-specific Setup Results}
\label{sec:subsec4.1}

\begin{figure*}[ht!]
    \centering
    \includegraphics[width=\textwidth]{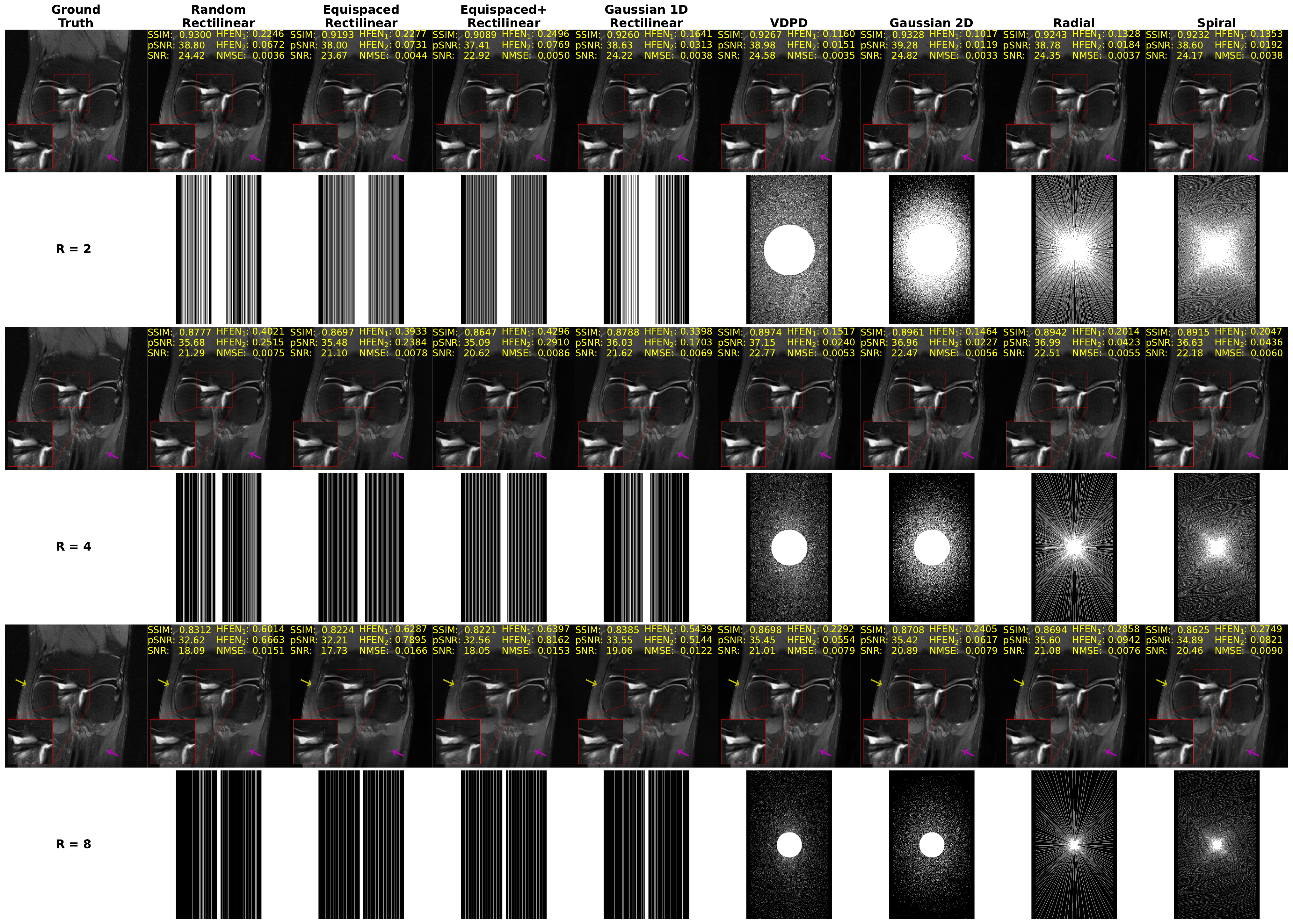}
    \caption{Representative reconstruction of a sample from the fastMRI knee test set obtained from the scheme-specific experiments. For the ground truth image the RSS method was applied on the fully sampled $k$-space. For each subsampling scheme, we retrospectively applied subsampling masks as shown in rows 2, 4, 6 (with accelerations factors of 2, 4, 8, respectively) onto the fully sampled $k$-space. Rows 1, 3, 5 illustrate the center-cropped RSS-reconstructed $k$-space output from each model. Quantitative metrics against the ground truth are inscribed on the top left of each reconstruction.}
    \label{fig:scheme-specific_knee_results}
\end{figure*}

\begin{figure*}[!ht]
    \centering
    \includegraphics[width=\textwidth]{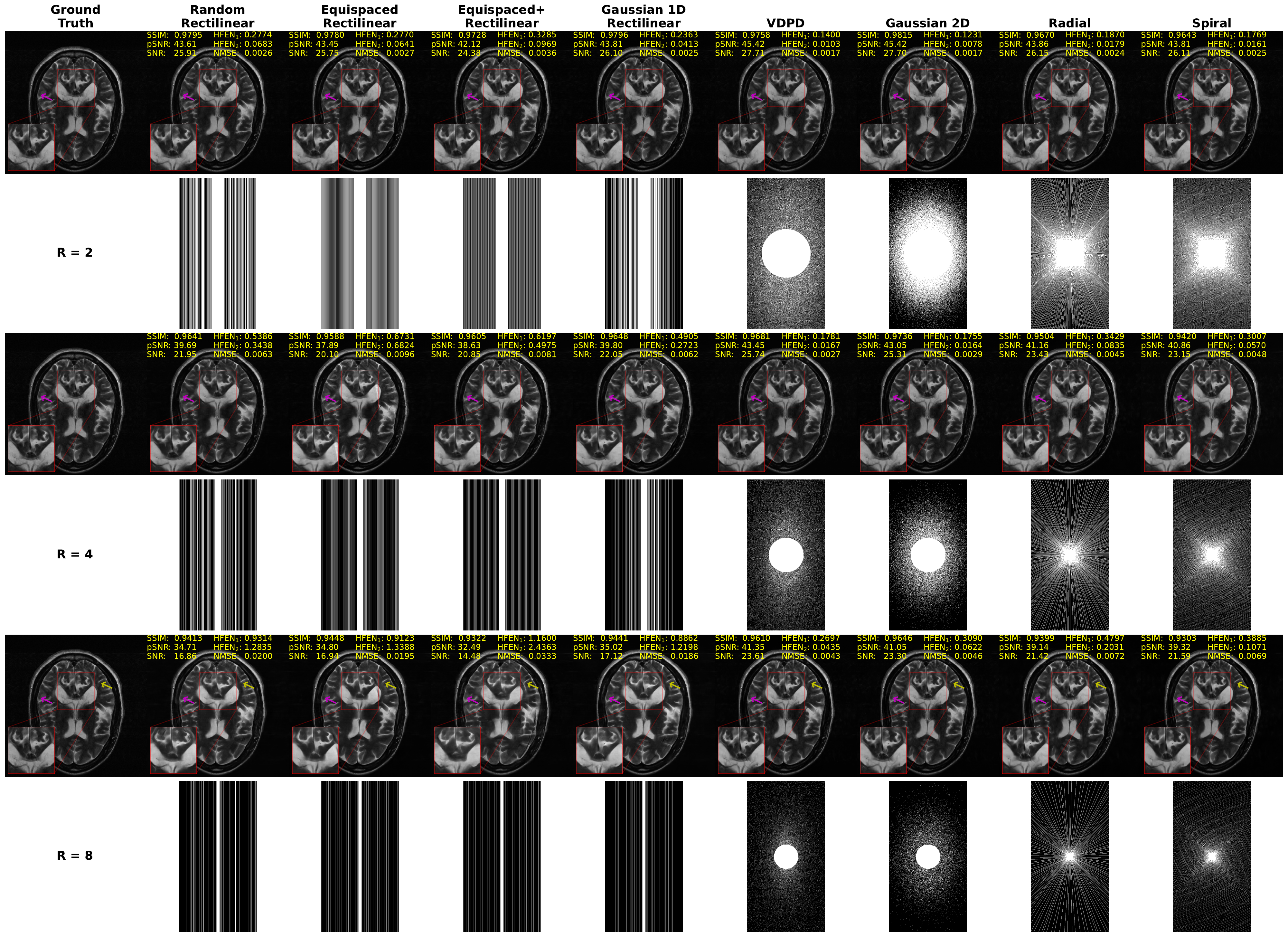}
    \caption{Representative reconstruction of a T2-weighted sample from the fastMRI brain test set obtained from the scheme-specific experiments. For the ground truth image the RSS method was applied on the fully sampled $k$-space. For each subsampling scheme, we retrospectively applied subsampling masks as shown in rows 2, 4, 6 (with accelerations factors of 2, 4, 8, respectively) onto the fully sampled $k$-space. Rows 1, 3, 5 illustrate the center-cropped RSS-reconstructed $k$-space output from each model. Quantitative metrics against the ground truth are inscribed on the top left of each reconstruction.}
    \label{fig:scheme-specific_brain_results}
\end{figure*}

\Figure{metrics} illustrates the quantitative metrics computed on the test sets in the form of violin-plots. The average metrics are reported in \Table{A2}. Additionally, for visual assessment, in \Figure{scheme-specific_knee_results}, \Figure{scheme-specific_brain_results}, and \Figure{A1} we present example reconstructions of a test sample from each dataset using all methods along with ground truths and the retrospective subsampling mask used in each scenario.


\subsubsection{Cartesian}
\label{sec:subse}
\paragraph{Rectilinear}
\Figure{metrics} indicates that all models trained on rectilinearly subsampled measurements achieved comparable results when evaluated on the test sets subsampled with the respective rectilinear schemes. Moreover, although Gaussian 1D subsampling outperformed the rest of the rectilinear schemes in the case of both fastMRI datasets, the opposite was noted in the case of the CC dataset. This is also visible by the example reconstructions in Figures \ref{fig:scheme-specific_knee_results}, \ref{fig:scheme-specific_brain_results}, and \Figure{A1},  especially for 8-fold acceleration.

\paragraph{Non-Rectilinear}
As \Figure{metrics} and \Table{A2} indicate, models trained on data subsampled with either VDPD or Gaussian 2D schemes were the best-performing models when evaluated on the respective data. Additionally, they produced reconstructions with the higher average SSIM, pSNR, and SNR and lower HFEN and NMSE values for all combinations of datasets and acceleration factors.

\subsubsection{Non-Cartesian}
As shown in \Figure{metrics} and \Table{A2}, models trained and evaluated on simulated non-Cartesian (radially and spirally) subsampled data yielded similar performance to Cartesian non-rectilinear schemes. In general, spiral schemes outperformed radial schemes for all combinations of datasets and acceleration factors, though the difference in performance was minor.

\subsubsection{Comparisons}
\label{subsubsec:4.1.3}
Considering \Figure{metrics}, models trained on non-rectilinearly subsampled data, produced reconstructions of higher fidelity when evaluated on the respective test sets in comparison to models trained and evaluated on rectilinearly subsampled data for all datasets and acceleration factors. Models trained with VDPD or Gaussian 2D -subsampled $k$-spaces obtained the best average quantitative evaluation results. 

As can be seen from Figures \ref{fig:scheme-specific_knee_results}, \ref{fig:scheme-specific_brain_results}, and \ref{fig:A1}, for high acceleration factors (4 or 8) models trained on rectilinear schemes reconstructed images with more errors and artifacts, compared to non-Cartesian or non-rectilinear schemes. However, for $R=2$ all models performed similarly.

\subsection{Multi-scheme Setup Results}
\label{sec:subsec4.2}

\begin{figure*}[!hbt]
    \centering
    \includegraphics[width=0.92\textwidth]{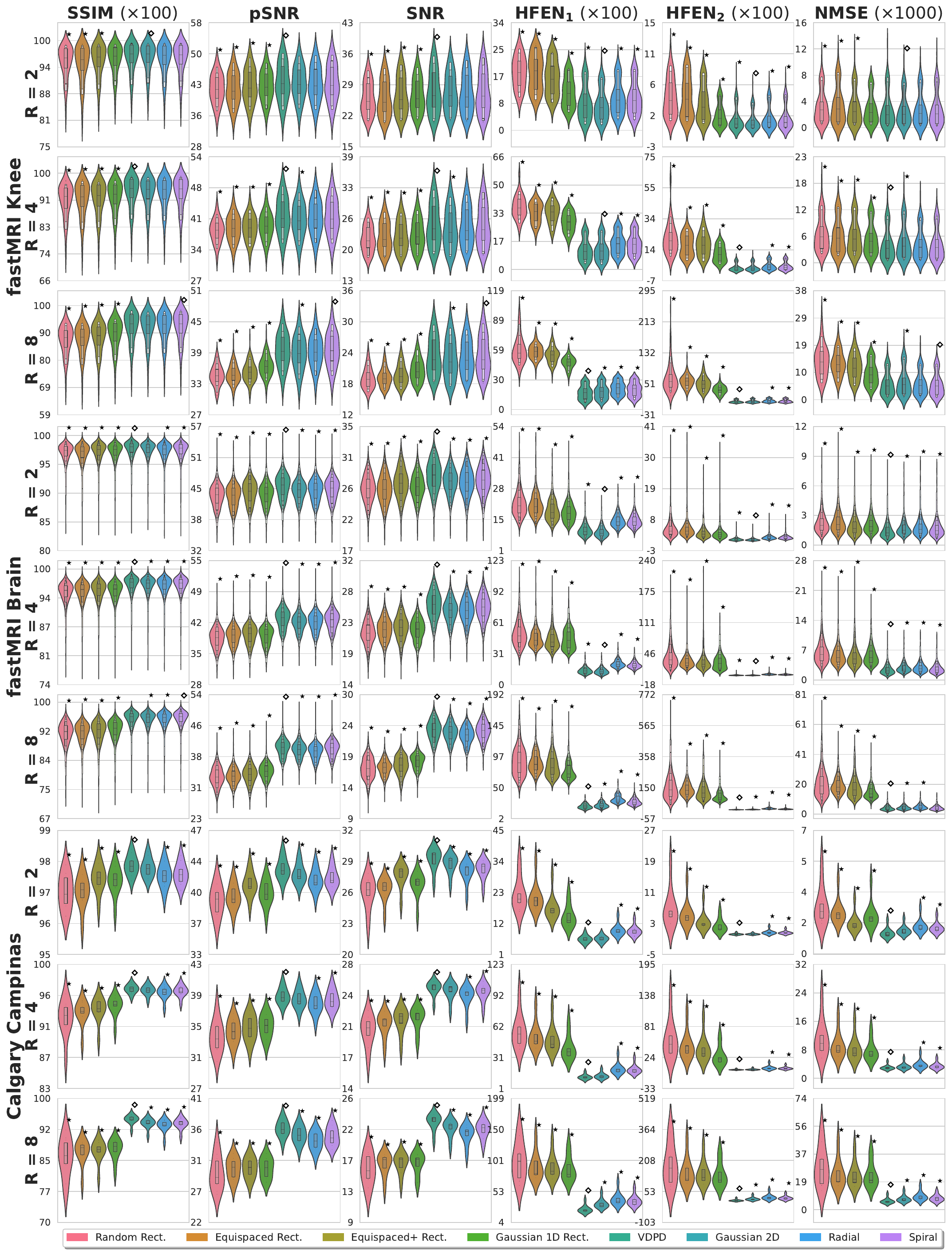}
    \caption{Multi-scheme experiments quantitative results on the test sets. A model was trained for each dataset on data subsampled by any of the presented schemes. For each dataset-acceleration-metric combination, pair-wise ASO significance tests were performed between the average best performing (denoted by $\diamond$) and the rest schemes. The symbol $\star$ indicates that the best scheme was found to be significantly better ($\epsilon_\text{min} < 0.5$). Average metrics are reported in \Table{A3}.}
    \label{fig:multi-scheme_metrics}
\end{figure*}

Figure \ref{fig:multi-scheme_metrics} reports the quantitative evaluation results on the test sets in the multi-scheme setup and \Table{A3} the corresponding average metrics. We observe that, similarly to the results of \Section{subsec4.1}, reconstructions of non-rectilinearly subsampled measurements produced better quantitative results compared to reconstructions of rectilinearly subsampled data. In addition, VDPD and Gaussian 2D-subsampled reconstructions were the highest performing on average.

\begin{table*}[!hbt]
\centering

\caption{Scheme-specific-Multi-scheme experiments performance average percentage difference. Percentages were acquired by averaging the per-case differences (as calculated and illustrated in \Figure{A2}) for all dataset-scheme type (rectilinear or non-rectilinear) combinations.}
\label{tab:univsmulti}
\resizebox{0.85\textwidth}{!}{%
\begin{tabular}{cccccccccccccc}
\hline
\multirow{2}{*}{Dataset}                                                    & \multirow{2}{*}{\begin{tabular}[c]{@{}c@{}}Acceleration\\ Factor\end{tabular}} & \multicolumn{6}{c}{Rectilinear}                                             & \multicolumn{6}{c}{Non-rectilinear}                                           \\ \cline{3-14} 
                                                                            &                                                                                & SSIM  & pSNR  & SNR   & HFEN$_{\text{L}_1}$ & HFEN$_{\text{L}_2}$ & NMSE    & SSIM   & pSNR   & SNR    & HFEN$_{\text{L}_1}$ & HFEN$_{\text{L}_2}$ & NMSE   \\ \hline
\multirow{3}{*}{\begin{tabular}[c]{@{}c@{}}fastMRI\\ Knee\end{tabular}}     & 2                                                                              & 0.1\% & 0.5\% & 0.7\% & -2.7\%              & -5.3\%              & -0.7\%  & -0.0\% & -0.3\% & -0.7\% & 4.4\%               & 7.1\%               & 5.6\%  \\
                                                                            & 4                                                                              & 0.3\% & 0.7\% & 0.9\% & -4.3\%              & -11.0\%             & -3.1\%  & -0.0\% & -0.2\% & -0.7\% & 2.2\%               & 4.8\%               & 4.2\%  \\
                                                                            & 8                                                                              & 0.7\% & 1.4\% & 2.1\% & -6.2\%              & -15.4\%             & -9.5\%  & 0.0\%  & -0.1\% & -0.5\% & -0.2\%              & 0.3\%               & 3.0\%  \\ \hline
\multirow{3}{*}{\begin{tabular}[c]{@{}c@{}}fastMRI\\ Brain\end{tabular}}    & 2                                                                              & 0.0\% & 0.3\% & 0.4\% & -0.0\%              & -0.6\%              & -2.6\%  & -0.0\% & -0.2\% & -0.3\% & 6.2\%               & 13.2\%              & 1.6\%  \\
                                                                            & 4                                                                              & 0.1\% & 0.3\% & 0.4\% & -0.5\%              & -1.9\%              & -2.7\%  & -0.0\% & -0.1\% & -0.0\% & 1.3\%               & 3.4\%               & 0.5\%  \\
                                                                            & 8                                                                              & 0.4\% & 1.2\% & 2.1\% & -4.2\%              & -9.8\%              & -8.5\%  & 0.0\%  & 0.1\%  & 0.3\%  & 2.6\%               & 11.3\%              & -0.9\% \\ \hline
\multirow{3}{*}{\begin{tabular}[c]{@{}c@{}}Calgary\\ Campinas\end{tabular}} & 2                                                                              & 0.3\% & 1.5\% & 2.2\% & -9.1\%              & -19.1\%             & -12.9\% & 0.0\%  & 0.1\%  & 0.1\%  & -1.2\%              & -4.7\%              & -0.8\% \\
                                                                            & 4                                                                              & 0.9\% & 2.2\% & 3.5\% & -10.6\%             & -20.1\%             & -14.8\% & 0.0\%  & -0.1\% & -0.1\% & 0.9\%               & 2.2\%               & 0.7\%  \\
                                                                            & 8                                                                              & 2.3\% & 3.1\% & 5.7\% & -12.2\%             & -22.1\%             & -17.8\% & -0.0\% & -0.2\% & -0.3\% & 3.1\%               & 6.0\%               & 2.0\%  \\ \hline
\end{tabular}%
}
\end{table*}

For further investigation, in \Figure{A2}, utilizing the results reported in \Figure{multi-scheme_metrics}, we calculated the per-case difference in evaluation metrics change for each pattern using as reference the results obtained in the scheme-specific setup (\Figure{metrics}). We also report the average differences for rectilinear and non-rectilinear patterns in \Table{univsmulti}. Interestingly, models trained in the multi-scheme setting exemplified superior performance when evaluated on measurements subsampled with rectilinear schemes compared against the models trained on individual schemes when evaluated on the same data. In particular, in the case of rectilinear schemes, noticeable improvements (SSIM/pSNR/SNR increase, HFEN/NMSE decrease) on the reconstruction performance were remarked for all datasets and acceleration factors, whilst for non-rectilinear patterns no change or minor deterioration was observed. 

\begin{figure*}[!hbt]
    \centering
    \includegraphics[width=\textwidth]{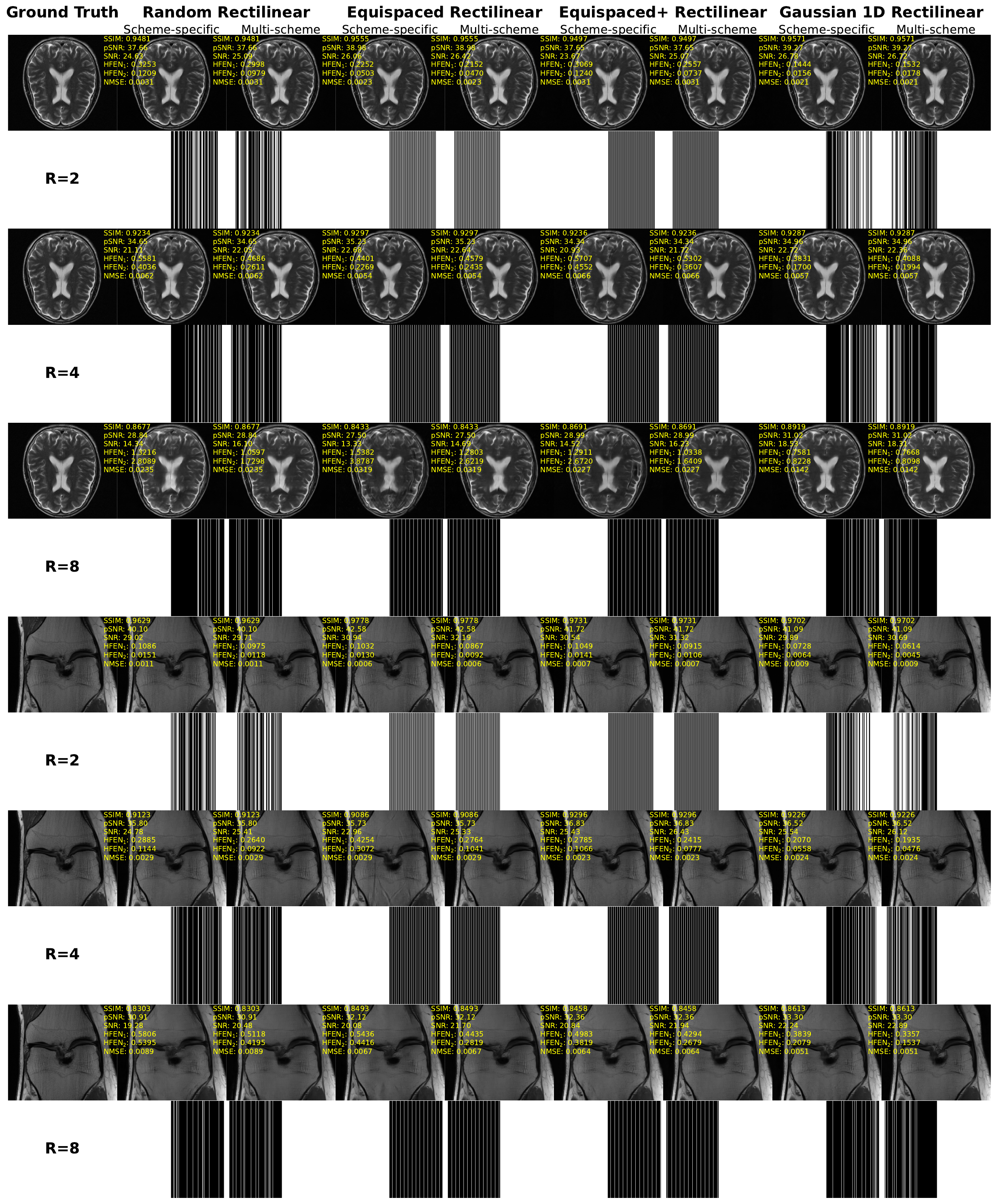}
    \caption{Scheme-specific vs multi-scheme setup visual comparison: Representative reconstructions of two samples from the fastMRI datasets. Measurements were subsampled with the four rectilinear schemes  for three acceleration factors.}
    \label{fig:univsmulti_fastmri}
\end{figure*}

In Figures \ref{fig:univsmulti_fastmri} and \ref{fig:A3} we show representative reconstructions predicted from rectilinearly-subsampled data obtained from both, the scheme-specific and multi-scheme setups. By visual investigation, we notice that the latter reconstructed images were more faithful.

\subsection{Robustness to Model Choice Results}
\label{sec:subsec4.3}
The quantitative metrics for our experiments using the Recurrent Inference Machine (RIM) are shown in \Figure{A4} and \Table{A4}. These results are on par with the findings of our original experiments, indicating that our conclusions remain valid regardless of the specific deep learning architecture employed.
\section{Discussion}
\label{sec:sec5}

In this work we investigated and compared various retrospective $k$-space subsampling patterns and we experimentally studied their effect on the quality of DL-based reconstructions. Since the data we utilized in our studies were Cartesian fully-sampled acquisitions, we retrospectively generated subsampling masks on the Cartesian grid as demonstrated in \Section{subsec3.1}, simulating prospective Cartesian or non-Cartesian accelerated acquisitions.  In particular, we generated four Cartesian rectilinear schemes (random, equispaced, equispaced with symmetry correction, Gaussian 1D), non-rectilinear Cartesian schemes (Variable Density Poisson-disk and Gaussian 2D), and simulated (non-Cartesian) radial and spiral schemes. 

Our experiments consisted of utilizing a state-of-the-art DL-based accelerated MRI reconstruction method - the Recurrent Variational Network. Although choosing the optimal DL-based MRI reconstruction algorithm was out of the scope of this project, it is important to note that other DL-based models do exist with similar performance. However, the RecurrentVarNet was selected due to its proven ability to produce high-quality reconstructions and its superior performance compared to other models in the field. Additionally, to demonstrate that the results were not model-dependent, the scheme-specific experiments on the CC dataset were repeated using the RIM model which produced on-par results.

Experiments were performed under two setups: scheme-specific and multi-scheme setups. In the scheme-specific settings, we trained individual models on data retrospectively subsampled with individual subsampling patterns applying various acceleration factors. Quantitative and qualitative results demonstrated that the models trained on conventional rectilinear schemes, compared to the other schemes, produced lower quality reconstructions with more artifacts especially for higher acceleration factors (4 or 8). This can be attributed to the fact that non-rectilinear Cartesian such as the VDPD or the Gaussian 2D, and non-Cartesian patterns such as the radial or spiral, allow for more incoherent sampling. This means that these schemes result in a more randomized and less correlated distribution of samples, reducing the dependence of the reconstruction quality on any specific pattern of missing data. Additionally, these schemes allow for center oversampling, which contains information such as contrast and the general shape of the reconstruction.

In the multi-scheme setup, unified models were trained on all data subsampled by all types of subsampling patterns. Results showed that non-rectilinearly subsampled data reconstructions were of superior quality compared to rectilinear reconstructions, aligning with the findings from the scheme-specific experiments. Despite this, the reconstruction performance for rectilinearly subsampled measurements improved notably compared to the scheme-specific setup (see \Table{univsmulti} and \Figure{A2}). Additionally, the violin-plots in \Figure{metrics} and \Figure{multi-scheme_metrics} show that in both setups for most dataset-metric-acceleration combinations quantitative results for the four rectilinear and the four non-rectilinear schemes formed similar metric distributions and were with similar numbers of outliers, suggesting that results were also case-dependent.

Even though comparing prospective acquisition speeds was out of the scope of this project, we provide a brief discussion. While the reconstruction performance of Cartesian rectilinearly subsampled data was inferior to Cartesian non-rectilinear, someone could argue about the trade-off between acquisition speed and quality, as MRI scanners can perform rectilinear sampling in fast acquisition times \cite{21391}, while strategies such as VDPD or Gaussian 2D can be slower due to physical limitations. For instance, in the prospective case these strategies may require large gradient switches in the MRI scanner due to the enlarged $k$-space spacing which can cause extended times due to hardware constraints, whilst in the retrospective settings, efficient algorithms are used to pick the samples. 

On the other hand, our results indicated that synthesized radial or spiral schemes on the Cartesian grid using the CIRCUS method provided similar performance to VDPD and Gaussian 2D schemes. In the original paper \cite{circus}, the authors state that their method improved sampling efficiency over VDPD while maintaining the reconstruction performance in the prospective case. 

However, prospective non-Cartesian sampling trajectories do not sample on a Cartesian grid, and therefore samples closer to the center are more densely placed, and more scattered far off. As a result of this non-uniformity, the application of the (inverse) non-uniform FFT (NUFFT) \cite{Fourmont_2003} is required, which includes a (inverse) gridding process \cite{Jackson_1991}, and possibly accumulating additional computation times. 

The main limitation of this study is the fact that experiments were performed retrospectively using fully sampled data. It is important to note that comparing Cartesian and non-Cartesian data acquired prospectively could introduce confounding factors such as different contrasts, potentially leading to unfair comparisons.  Nevertheless, the retrospective approach allowed us to directly compare the performance of DL-based reconstructions on various subsampling patterns, thus enabling a clearer understanding of their relative strengths and limitations.

Future studies could explore repeating our experiments using prospectively subsampled data with the different schemes we employed in this study, including non-Cartesian acquisition, in order to determine the extend to which our findings hold true in prospective settings. Additionally, an alternative approach would be to generate prospective non-Cartesian acquisitions from the available Cartesian fully sampled data.  To that end it will be necessary to apply one of the methods discussed in the previous paragraph to handle non-Cartesian data in order to incorporate it into our deep learning pipeline and match the multi-scheme framework used in this study. 

Another limitation worth considering is that while in our experiments we retrospectively acquired the ACS region and subsampled measurements from the same fully-sampled k-space, this may not be the case in the prospective scenario, as the ACS region may be sampled at different timings. This variation in the sampling of the ACS region may affect the contrast of the reconstructed image, and therefore warrants further investigation.

\section{Conclusion}
\label{sec:sec6}

In summary, our study sheds light on the impact of $k$-space subsampling patterns on DL-based MRI reconstructions. By comparing rectilinear and non-rectilinear Cartesian patterns, as well as pseudo-non-Cartesian trajectories, we demonstrated that DL-based reconstructions perform more robustly and accurately trained with non-rectilinearly subsampled data, allowing for higher acceleration factors. The evaluations conducted using the six quality metrics (SSIM, pSNR, SNR, HFEN$_1$, HFEN$_2$, NMSE) consistently showed that non-rectilinear schemes outperformed rectilinear schemes in terms of reconstruction quality. Our work provides a foundation for future research in exploring DL-based MRI reconstructions with different sampling schemes and encourages further investigation into advanced non-Cartesian trajectories.

% \input{chapters/supporting}

% %% The Appendices part is started with the command \appendix;
% %% appendix sections are then done as normal sections

%%
\bibliographystyle{elsarticle-num} 
\bibliography{report}

\onecolumn
\appendix

\setcounter{figure}{0}
\setcounter{table}{0}

\newpage
\section{Supplementary Material}
\label{sec:appendix1}

In this section, we provide additional figures and tables that support the findings and analysis presented in the main paper.

\subsection{Acquisition Parameters and Experiment Splits}
\setlength{\tabcolsep}{0.8pt}
{\renewcommand{\arraystretch}{1.7}

\begin{table}[!hbt]
\caption{Acquisition parameters and experiment splits per dataset used in our experiments.}
\centering
\resizebox{\textwidth}{!}{%
\begin{tabular}{ccccc}
\hline
\multicolumn{2}{c}{\textbf{Dataset}}                                                                                      & \textbf{FastMRI Knee}                                                               & \textbf{FastMRI Brain}                                                                & \textbf{Calgary Campinas}                                                      \\ \hline
\multicolumn{2}{c}{\textbf{Field Strength}}                                                                               & 1.5T \& 3.0T                                                                        & 1.5T \& 3.0T                                                                          & 3.0T                                                                           \\
\multicolumn{2}{c}{\textbf{Sequence}}                                                                                     & \begin{tabular}[c]{@{}c@{}}PD $^1$ with and \\ without fat suppression\end{tabular} & \begin{tabular}[c]{@{}c@{}}T1-w $^2$, T1-w post contrast, \\ T2-w, FLAIR\end{tabular} & T1-w                                                                           \\
\multicolumn{2}{c}{\textbf{Subjects}}                                                                                     & \begin{tabular}[c]{@{}c@{}}Healthy or \\ Abnormality present\end{tabular}           & \begin{tabular}[c]{@{}c@{}}Healthy or \\ Pathology present\end{tabular}               & \begin{tabular}[c]{@{}c@{}}Presumed \\ healthy\end{tabular}                    \\
\multicolumn{2}{c}{\textbf{Acquisition}}                                                                                  & Cartesian                                                                           & Cartesian                                                                             & Cartesian                                                                      \\
\multicolumn{2}{c}{\textbf{Fully Sampled}}                                                                                & Yes                                                                                 & Yes                                                                                   & Yes                                                                            \\
\multicolumn{2}{c}{\textbf{\begin{tabular}[c]{@{}c@{}}Subsampling \\ Directions\end{tabular}}}                            & \begin{tabular}[c]{@{}c@{}}Phase and slice \\ encoding directions\end{tabular}      & \begin{tabular}[c]{@{}c@{}}Phase and slice \\ encoding directions\end{tabular}        & \begin{tabular}[c]{@{}c@{}}Phase and slice \\ encoding directions\end{tabular} \\
\multicolumn{2}{c}{\textbf{No. Coils}}                                                                                    & 15                                                                                  & 2 - 24                                                                                & 12                                                                             \\
\multicolumn{2}{c}{\textbf{No. Volumes}}                                                                                  & 1172                                                                                & 5846                                                                                  & 67                                                                             \\
\multicolumn{2}{c}{\textbf{No. Slices}}                                                                                   & 41877                                                                               & 92574                                                                                 & 10452                                                                          \\
\multirow{3}{*}{\textbf{\begin{tabular}[c]{@{}c@{}}Split size \\ (No. volumes/ \\ No. slices)\end{tabular}}} & Training   & 973 / 34742                                                                         & 4469 / 70748                                                                          & 40 / 6240                                                                      \\
                                                                                                             & Validation & 100 / 3587                                                                          & 686 / 10880                                                                           & 14 / 2184                                                                      \\
                                                                                                             & Test       & 99 / 3548                                                                           & 691 / 10946                                                                           & 13 / 2028                                                                      \\ \hline
\end{tabular}%
}
\begin{tablenotes}%%[\textheight]
\item[$^{1}$] Proton Density
\item[$^{2}$] Weighted
\end{tablenotes}
\label{tab:A1}
\end{table}
}
\newpage

\subsection{Scheme-Specific Setup Experiments Results}
\begin{figure}[!hbt]
\setlength{\linewidth}{\textwidth}
  \setlength{\hsize}{\textwidth}
  \centering\includegraphics[width=1\textwidth]{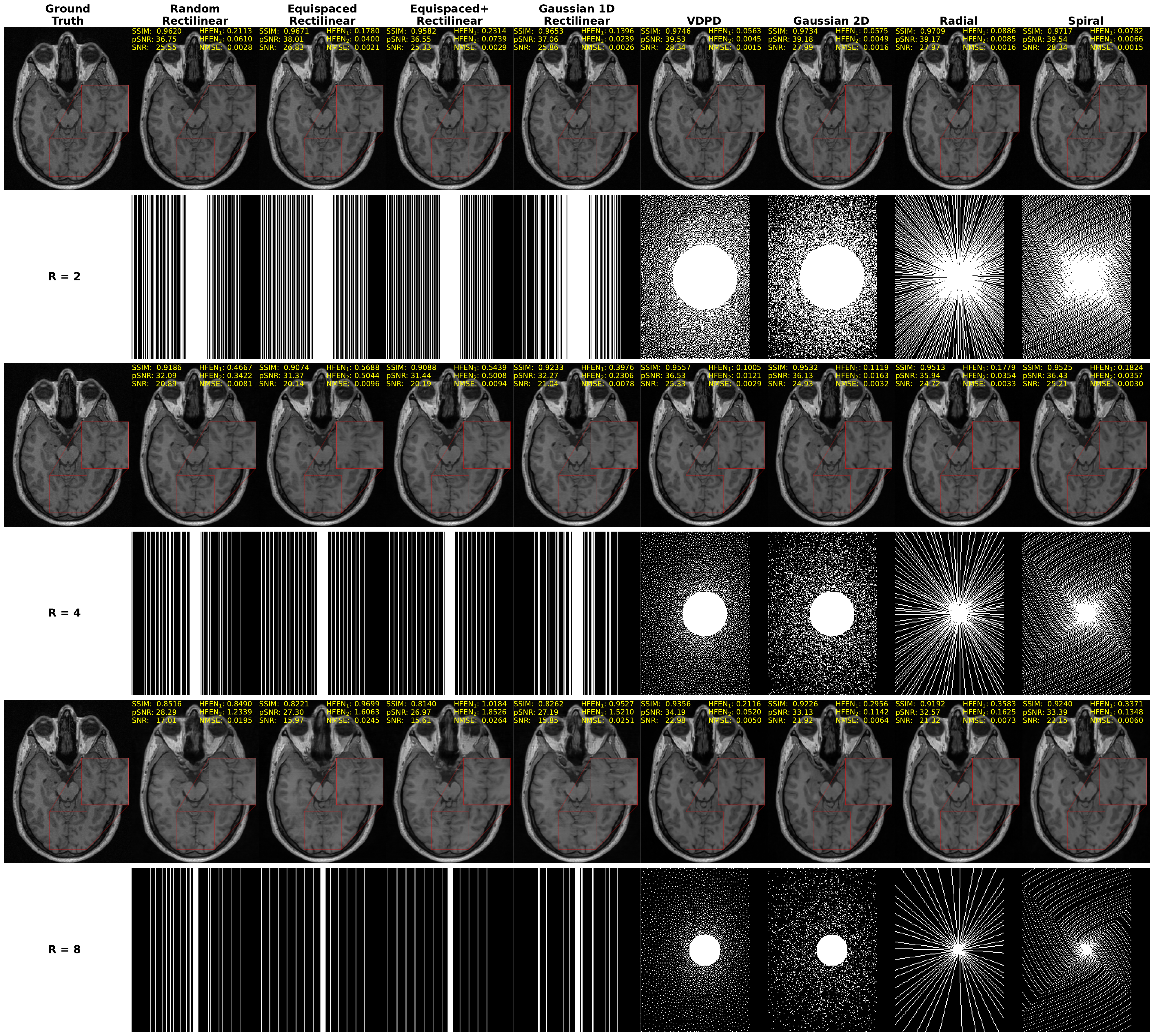}
    \caption{Representative reconstruction of a sample from the Calgary Campinas test set obtained from the scheme-specific experiments. For the ground truth image the RSS method was applied on the fully sampled $k$-space. For each subsampling scheme, we retrospectively applied subsampling masks as shown in rows 2, 4, 6 (with accelerations factors of 2, 4, 8, respectively) onto the fully sampled $k$-space. Rows 1, 3, 5 illustrate the RSS reconstructed $k$-space output from each model. Quantitative metrics against the ground truth are inscribed on the top left of each reconstruction. Note that the right 15\% portion of the $k$-space was not collected and therefore subsampling masks were zero-filled accordingly.}
        \label{fig:A1}
\end{figure}
\newpage

\setlength{\tabcolsep}{0.8pt}
{\renewcommand{\arraystretch}{1.7}

 \begin{table*}[!hbt]
 \centering
  \caption{Scheme-specific experiments average results on the test sets. Bold numbers indicate the best across subsampling schemes.}
 \resizebox{\textwidth}{!}{
\begin{tabular}{cccccccccccccccccccc}
\hline
\multirow{3}{*}{Dataset}                                                    & \multirow{3}{*}{\begin{tabular}[c]{@{}c@{}}Subsampling\\ Scheme\end{tabular}} & \multicolumn{18}{c}{Acceleration Factor ($R$)}                                                                                                                                                                                                                                                            \\ \cline{3-20} 
                                                                            &                                                                               & \multicolumn{6}{c}{2}                                                                             & \multicolumn{6}{c}{4}                                                                             & \multicolumn{6}{c}{8}                                                                             \\ \cline{3-20} 
                                                                            &                                                                               & SSIM           & pSNR          & SNR           & HFEN$_{1}$   & HFEN$_{2}$   & NMSE           & SSIM           & pSNR          & SNR           & HFEN$_{1}$   & HFEN$_{2}$   & NMSE           & SSIM           & pSNR          & SNR           & HFEN$_{1}$   & HFEN$_{2}$   & NMSE           \\ \hline
\multirow{8}{*}{\begin{tabular}[c]{@{}c@{}}FastMRI\\ Knee\end{tabular}}     & Random Rect.                                                                  & 0.948          & 41.8          & 25.7          & 0.186          & 0.049          & 0.003          & 0.910          & 38.0          & 22.1          & 0.384          & 0.240          & 0.007          & 0.864          & 34.1          & 18.4          & 0.633          & 0.756          & 0.015          \\
                                                                            & Equispaced Rect.                                                              & 0.945          & 41.8          & 25.7          & 0.184          & 0.048          & 0.003          & 0.913          & 38.5          & 22.6          & 0.350          & 0.192          & 0.006          & 0.866          & 34.2          & 18.6          & 0.613          & 0.675          & 0.014          \\
                                                                            & Equispaced+ Rect.                                                             & 0.951          & 42.5          & 26.4          & 0.167          & 0.039          & 0.003          & 0.914          & 38.5          & 22.6          & 0.352          & 0.196          & 0.006          & 0.871          & 34.7          & 19.1          & 0.580          & 0.600          & 0.013          \\
                                                                            & Gaussian  1D                                                                  & 0.954          & 42.5          & 26.4          & 0.121          & 0.020          & 0.003          & 0.920          & 39.1          & 23.1          & 0.288          & 0.122          & 0.005          & 0.885          & 35.9          & 20.2          & 0.498          & 0.411          & 0.010          \\
                                                                            & VDPD                                                                          & \textbf{0.959} & \textbf{43.9} & \textbf{27.7} & 0.082          & 0.013          & \textbf{0.002} & \textbf{0.938} & \textbf{41.6} & \textbf{25.5} & 0.113          & \textbf{0.020} & \textbf{0.004} & 0.918          & \textbf{39.6} & \textbf{23.4} & \textbf{0.170} & \textbf{0.039} & \textbf{0.005} \\
                                                                            & Gaussian  2D                                                                  & 0.959          & 43.5          & 27.3          & \textbf{0.076} & \textbf{0.011} & 0.003          & 0.935          & 40.9          & 24.7          & \textbf{0.113} & 0.021          & 0.004          & 0.916          & 39.2          & 23.0          & 0.181          & 0.043          & 0.006          \\
                                                                            & Radial                                                                        & 0.956          & 43.3          & 27.1          & 0.096          & 0.015          & 0.003          & 0.935          & 41.1          & 24.9          & 0.149          & 0.030          & 0.004          & 0.916          & 39.3          & 23.1          & 0.211          & 0.058          & 0.005          \\
                                                                            & Spiral                                                                        & 0.957          & 43.5          & 27.3          & 0.096          & 0.015          & 0.003          & 0.937          & 41.3          & 25.0          & 0.151          & 0.031          & 0.004          & \textbf{0.919} & 39.5          & 23.3          & 0.205          & 0.055          & 0.005          \\ \hline
\multirow{8}{*}{\begin{tabular}[c]{@{}c@{}}FastMRI\\ Brain\end{tabular}}    & Random Rect.                                                                  & 0.970          & 42.8          & 26.1          & 0.223          & 0.052          & 0.002          & 0.944          & 37.8          & 21.4          & 0.539          & 0.466          & 0.007          & 0.903          & 32.4          & 16.3          & 1.000          & 1.999          & 0.026          \\
                                                                            & Equispaced Rect.                                                              & 0.969          & 43.0          & 26.2          & 0.202          & 0.040          & 0.002          & 0.949          & 38.9          & 22.4          & 0.441          & 0.280          & 0.006          & 0.911          & 33.3          & 17.2          & 0.887          & 1.443          & 0.020          \\
                                                                            & Equispaced+ Rect.                                                             & 0.972          & 43.7          & 27.0          & 0.187          & 0.033          & 0.002          & 0.949          & 38.9          & 22.5          & 0.449          & 0.297          & 0.006          & 0.912          & 33.6          & 17.4          & 0.868          & 1.394          & 0.019          \\
                                                                            & Gaussian  1D                                                                  & 0.972          & 43.4          & 26.7          & 0.174          & 0.027          & 0.002          & 0.950          & 38.9          & 22.3          & 0.442          & 0.284          & 0.006          & 0.920          & 34.4          & 18.1          & 0.765          & 1.012          & 0.016          \\
                                                                            & VDPD                                                                          & \textbf{0.977} & \textbf{45.2} & \textbf{28.3} & 0.098          & 0.008          & \textbf{0.001} & \textbf{0.966} & \textbf{42.8} & \textbf{25.9} & 0.134          & 0.014          & \textbf{0.002} & \textbf{0.954} & \textbf{40.5} & \textbf{23.7} & \textbf{0.199} & \textbf{0.030} & \textbf{0.004} \\
                                                                            & Gaussian  2D                                                                  & 0.976          & 44.5          & 27.6          & \textbf{0.090} & \textbf{0.007} & 0.002          & 0.965          & 42.0          & 25.2          & \textbf{0.131} & \textbf{0.013} & 0.003          & 0.953          & 40.0          & 23.2          & 0.237          & 0.046          & 0.004          \\
                                                                            & Radial                                                                        & 0.973          & 44.1          & 27.2          & 0.133          & 0.013          & 0.002          & 0.962          & 41.7          & 24.9          & 0.207          & 0.035          & 0.003          & 0.950          & 39.3          & 22.6          & 0.315          & 0.105          & 0.005          \\
                                                                            & Spiral                                                                        & 0.975          & 44.5          & 27.6          & 0.128          & 0.012          & 0.002          & 0.964          & 42.2          & 25.3          & 0.198          & 0.031          & 0.003          & 0.953          & 40.1          & 23.3          & 0.259          & 0.058          & 0.004          \\ \hline
\multirow{8}{*}{\begin{tabular}[c]{@{}c@{}}Calgary\\ Campinas\end{tabular}} & Random Rect.                                                                  & 0.967          & 38.5          & 25.2          & 0.252          & 0.091          & 0.003          & 0.913          & 32.5          & 19.5          & 0.653          & 0.738          & 0.014          & 0.834          & 28.4          & 15.2          & 1.150          & 2.375          & 0.035          \\
                                                                            & Equispaced Rect.                                                              & 0.968          & 39.1          & 25.8          & 0.229          & 0.070          & 0.003          & 0.925          & 33.4          & 20.5          & 0.573          & 0.553          & 0.011          & 0.853          & 29.2          & 16.2          & 1.015          & 1.829          & 0.028          \\
                                                                            & Equispaced+ Rect.                                                             & 0.974          & 40.5          & 27.0          & 0.194          & 0.046          & 0.002          & 0.928          & 33.8          & 20.8          & 0.553          & 0.507          & 0.010          & 0.851          & 29.2          & 16.2          & 1.029          & 1.837          & 0.028          \\
                                                                            & Gaussian  1D                                                                  & 0.973          & 39.6          & 26.3          & 0.160          & 0.033          & 0.003          & 0.935          & 34.3          & 21.2          & 0.450          & 0.329          & 0.009          & 0.854          & 29.1          & 16.1          & 1.018          & 1.829          & 0.029          \\
                                                                            & VDPD                                                                          & \textbf{0.981} & \textbf{42.8} & \textbf{29.2} & \textbf{0.065} & \textbf{0.005} & \textbf{0.001} & \textbf{0.966} & \textbf{39.0} & \textbf{25.5} & \textbf{0.118} & \textbf{0.016} & \textbf{0.003} & \textbf{0.949} & \textbf{36.3} & \textbf{23.0} & \textbf{0.231} & \textbf{0.067} & \textbf{0.005} \\
                                                                            & Gaussian  2D                                                                  & 0.980          & 42.3          & 28.7          & 0.066          & 0.005          & 0.001          & 0.963          & 38.5          & 25.1          & 0.130          & 0.020          & 0.003          & 0.940          & 35.2          & 22.0          & 0.315          & 0.136          & 0.007          \\
                                                                            & Radial                                                                        & 0.976          & 41.4          & 27.8          & 0.102          & 0.011          & 0.002          & 0.960          & 37.8          & 24.4          & 0.200          & 0.048          & 0.004          & 0.933          & 34.2          & 21.1          & 0.390          & 0.225          & 0.009          \\
                                                                            & Spiral                                                                        & 0.976          & 41.5          & 27.9          & 0.096          & 0.010          & 0.002          & 0.960          & 38.0          & 24.6          & 0.198          & 0.045          & 0.004          & 0.933          & 34.5          & 21.4          & 0.389          & 0.209          & 0.008          \\ \hline
\end{tabular}%
 }
 \label{tab:A2}
 \end{table*}
}
\newpage

\subsection{Multi-Scheme Setup Experiments Results}

\setlength{\tabcolsep}{0.8pt}
{\renewcommand{\arraystretch}{1.6}
\begin{table*}[!hbt]
\centering
\caption{Multi-scheme experiments average results on the test sets. Bold numbers indicate the best across subsampling schemes.}
\resizebox{\textwidth}{!}{%
\begin{tabular}{cccccccccccccccccccc}
\hline
\multirow{3}{*}{Dataset}                                                    & \multirow{3}{*}{\begin{tabular}[c]{@{}c@{}}Subsampling\\ Scheme\end{tabular}} & \multicolumn{18}{c}{Acceleration Factor ($R$)}                                                                                                                                                                                                                                                            \\ \cline{3-20} 
                                                                            &                                                                               & \multicolumn{6}{c}{2}                                                                             & \multicolumn{6}{c}{4}                                                                             & \multicolumn{6}{c}{8}                                                                             \\ \cline{3-20} 
                                                                            &                                                                               & SSIM           & pSNR          & SNR           & HFEN$_{1}$   & HFEN$_{2}$   & NMSE           & SSIM           & pSNR          & SNR           & HFEN$_{1}$   & HFEN$_{2}$   & NMSE           & SSIM           & pSNR          & SNR           & HFEN$_{1}$   & HFEN$_{2}$   & NMSE           \\ \hline
\multirow{8}{*}{\begin{tabular}[c]{@{}c@{}}FastMRI\\ Knee\end{tabular}}     & Random Rect.                                                                  & 0.949          & 42.0          & 25.9          & 0.181          & 0.045          & 0.003          & 0.912          & 38.3          & 22.3          & 0.368          & 0.213          & 0.006          & 0.870          & 34.6          & 18.8          & 0.596          & 0.650          & 0.013          \\
                                                                            & Equispaced Rect.                                                              & 0.946          & 42.0          & 25.9          & 0.178          & 0.045          & 0.003          & 0.915          & 38.8          & 22.8          & 0.334          & 0.169          & 0.006          & 0.872          & 34.8          & 19.0          & 0.576          & 0.571          & 0.013          \\
                                                                            & Equispaced+ Rect.                                                             & 0.952          & 42.7          & 26.6          & 0.162          & 0.037          & 0.003          & 0.917          & 38.8          & 22.9          & 0.335          & 0.173          & 0.006          & 0.878          & 35.2          & 19.5          & 0.543          & 0.503          & 0.012          \\
                                                                            & Gaussian  1D                                                                  & 0.955          & 42.7          & 26.5          & 0.118          & 0.020          & 0.003          & 0.921          & 39.3          & 23.2          & 0.278          & 0.112          & 0.005          & 0.889          & 36.4          & 20.5          & 0.466          & 0.345          & 0.009          \\
                                                                            & VDPD                                                                          & 0.959          & \textbf{43.6} & \textbf{27.4} & 0.086          & 0.014          & 0.003          & \textbf{0.938} & \textbf{41.4} & \textbf{25.2} & 0.116          & \textbf{0.022} & 0.004          & 0.918          & 39.5          & 23.3          & \textbf{0.169} & \textbf{0.039} & 0.005          \\
                                                                            & Gaussian  2D                                                                  & \textbf{0.959} & 43.4          & 27.1          & \textbf{0.079} & \textbf{0.012} & \textbf{0.003} & 0.934          & 40.6          & 24.3          & \textbf{0.116} & 0.022          & 0.005          & 0.916          & 39.0          & 22.7          & 0.181          & 0.044          & 0.006          \\
                                                                            & Radial                                                                        & 0.955          & 43.3          & 27.0          & 0.101          & 0.016          & 0.003          & 0.935          & 41.1          & 24.9          & 0.153          & 0.032          & 0.004          & 0.917          & 39.3          & 23.1          & 0.213          & 0.059          & 0.006          \\
                                                                            & Spiral                                                                        & 0.957          & 43.4          & 27.2          & 0.100          & 0.016          & 0.003          & 0.937          & 41.3          & 25.0          & 0.150          & 0.030          & 0.004          & \textbf{0.920} & \textbf{39.6} & \textbf{23.4} & 0.201          & 0.053          & \textbf{0.005} \\ \hline
\multirow{8}{*}{\begin{tabular}[c]{@{}c@{}}FastMRI\\ Brain\end{tabular}}    & Random Rect.                                                                  & 0.971          & 43.1          & 26.3          & 0.213          & 0.047          & 0.002          & 0.947          & 38.3          & 21.8          & 0.504          & 0.395          & 0.007          & 0.912          & 33.3          & 17.2          & 0.899          & 1.562          & 0.021          \\
                                                                            & Equispaced Rect.                                                              & 0.969          & 43.0          & 26.3          & 0.209          & 0.043          & 0.002          & 0.949          & 38.8          & 22.3          & 0.457          & 0.303          & 0.006          & 0.913          & 33.4          & 17.3          & 0.876          & 1.403          & 0.019          \\
                                                                            & Equispaced+ Rect.                                                             & 0.973          & 43.9          & 27.1          & 0.185          & 0.031          & 0.002          & 0.950          & 39.1          & 22.6          & 0.445          & 0.291          & 0.006          & 0.915          & 33.9          & 17.7          & 0.840          & 1.290          & 0.018          \\
                                                                            & Gaussian  1D                                                                  & 0.972          & 43.4          & 26.6          & 0.178          & 0.028          & 0.002          & 0.949          & 38.8          & 22.3          & 0.450          & 0.290          & 0.006          & 0.922          & 34.6          & 18.3          & 0.748          & 0.939          & 0.015          \\
                                                                            & VDPD                                                                          & \textbf{0.977} & \textbf{45.1} & \textbf{28.2} & 0.101          & 0.009          & \textbf{0.001} & \textbf{0.966} & \textbf{42.6} & \textbf{25.8} & 0.140          & 0.015          & \textbf{0.003} & 0.953          & \textbf{40.4} & \textbf{23.6} & \textbf{0.208} & \textbf{0.034} & \textbf{0.004} \\
                                                                            & Gaussian  2D                                                                  & 0.976          & 44.2          & 27.4          & \textbf{0.093} & \textbf{0.007} & 0.002          & 0.964          & 41.8          & 25.0          & \textbf{0.134} & \textbf{0.014} & 0.003          & 0.952          & 40.0          & 23.2          & 0.229          & 0.043          & 0.004          \\
                                                                            & Radial                                                                        & 0.973          & 44.1          & 27.3          & 0.142          & 0.015          & 0.002          & 0.962          & 41.8          & 25.0          & 0.208          & 0.036          & 0.003          & 0.952          & 39.4          & 22.8          & 0.323          & 0.118          & 0.005          \\
                                                                            & Spiral                                                                        & 0.974          & 44.4          & 27.6          & 0.141          & 0.016          & 0.002          & 0.964          & 42.3          & 25.5          & 0.195          & 0.030          & 0.003          & \textbf{0.954} & 40.3          & 23.5          & 0.276          & 0.073          & 0.004          \\ \hline
\multirow{8}{*}{\begin{tabular}[c]{@{}c@{}}Calgary\\ Campinas\end{tabular}} & Random Rect.                                                                  & 0.970          & 39.3          & 26.0          & 0.223          & 0.071          & 0.003          & 0.925          & 33.4          & 20.4          & 0.569          & 0.567          & 0.011          & 0.856          & 29.4          & 16.2          & 0.998          & 1.836          & 0.029          \\
                                                                            & Equispaced Rect.                                                              & 0.971          & 39.7          & 26.3          & 0.211          & 0.057          & 0.003          & 0.932          & 34.1          & 21.1          & 0.519          & 0.452          & 0.009          & 0.868          & 30.0          & 16.9          & 0.915          & 1.489          & 0.024          \\
                                                                            & Equispaced+ Rect.                                                             & 0.976          & 41.0          & 27.6          & 0.177          & 0.037          & 0.002          & 0.937          & 34.6          & 21.6          & 0.492          & 0.400          & 0.008          & 0.870          & 30.1          & 17.1          & 0.904          & 1.433          & 0.023          \\
                                                                            & Gaussian  1D                                                                  & 0.975          & 40.1          & 26.7          & 0.147          & 0.027          & 0.002          & 0.941          & 34.8          & 21.8          & 0.408          & 0.270          & 0.008          & 0.875          & 30.0          & 17.0          & 0.881          & 1.373          & 0.023          \\
                                                                            & VDPD                                                                          & \textbf{0.981} & \textbf{42.7} & \textbf{29.1} & \textbf{0.066} & \textbf{0.005} & \textbf{0.001} & \textbf{0.965} & \textbf{38.8} & \textbf{25.3} & \textbf{0.121} & \textbf{0.017} & \textbf{0.003} & \textbf{0.947} & \textbf{36.0} & \textbf{22.7} & \textbf{0.244} & \textbf{0.075} & \textbf{0.006} \\
                                                                            & Gaussian  2D                                                                  & 0.980          & 42.2          & 28.6          & 0.067          & 0.005          & 0.001          & 0.963          & 38.4          & 25.0          & 0.131          & 0.021          & 0.003          & 0.938          & 35.0          & 21.8          & 0.332          & 0.152          & 0.007          \\
                                                                            & Radial                                                                        & 0.977          & 41.5          & 27.9          & 0.097          & 0.010          & 0.002          & 0.960          & 37.8          & 24.4          & 0.204          & 0.051          & 0.004          & 0.932          & 34.1          & 21.0          & 0.407          & 0.242          & 0.009          \\
                                                                            & Spiral                                                                        & 0.977          & 41.7          & 28.1          & 0.094          & 0.009          & 0.002          & 0.961          & 38.3          & 24.9          & 0.194          & 0.043          & 0.003          & 0.936          & 34.8          & 21.7          & 0.378          & 0.194          & 0.008          \\ \hline
\end{tabular}%
}
 \label{tab:A3}
\end{table*}
} % arraystrech

\subsection{Comparison of Scheme-Specific Setup vs Multi-Scheme Setup}

\begin{figure*}[!hbt]
    \centering
    \includegraphics[width=0.92\textwidth]{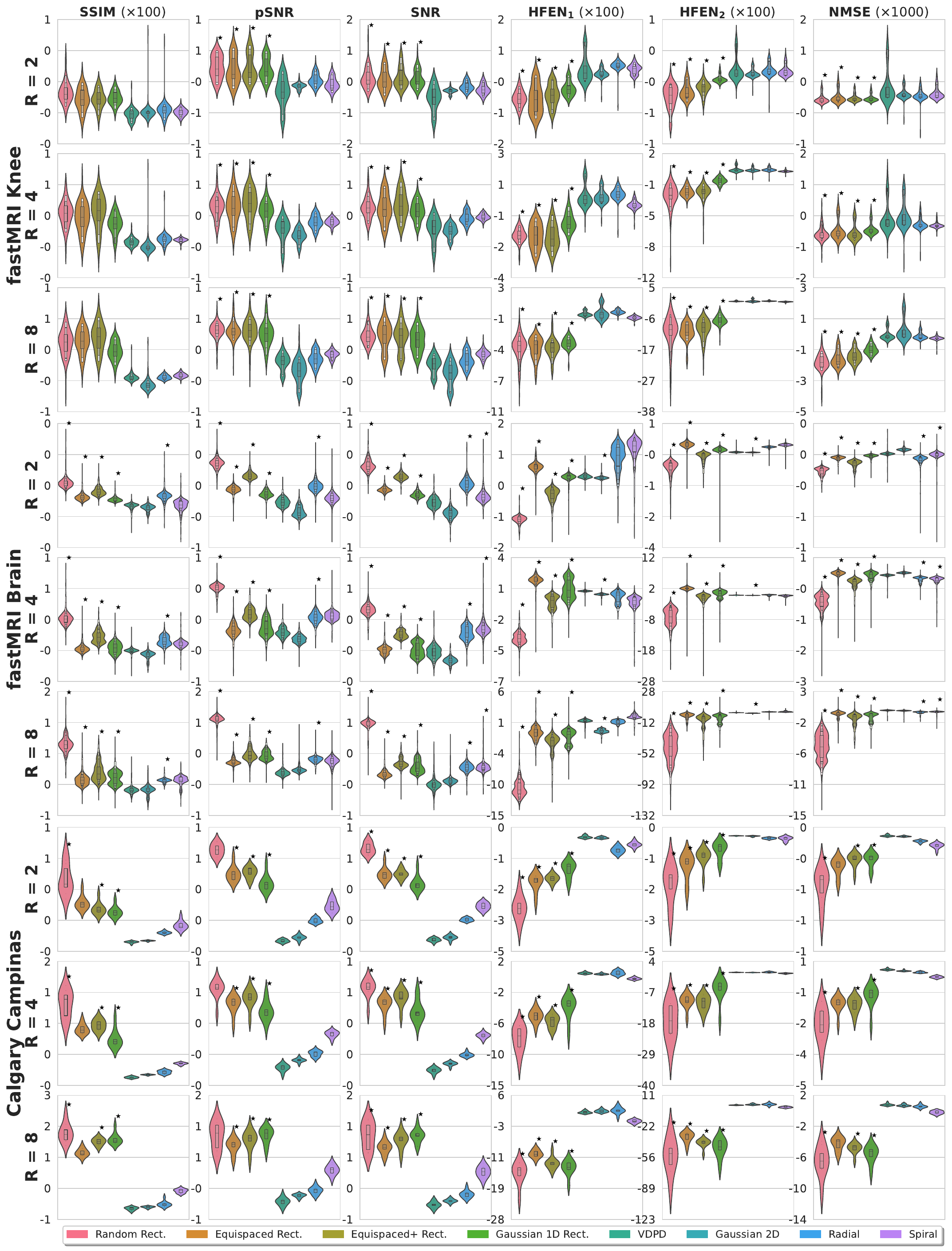}
    \caption{Comparison of Scheme-Specific Setup vs Multi-Scheme Setup: Quantitative Results.  The symbol $\star$ indicates a significant performance difference ($\epsilon_\text{min} < 0.5$).}
     \label{fig:A2}
\end{figure*}

In \Figure{A2}, we present a comparison between the Scheme-Specific Setup and Multi-Scheme Setup in terms of quantitative results. Violin plots are employed to visualize the variations in evaluation metrics between the two setups for each dataset. The Scheme-Specific Setup involves training a model on data subsampled using a single subsampling scheme, while the Multi-Scheme Setup trains a model on data subsampled using all available subsampling schemes.

To establish the necessary context, we define the relevant variables as follows: $\mathbf{y}$ represents the fully-sampled $k$-space data, and $\Tilde{\mathbf{y}}^{\text{S}}$ denotes the $k$-space data subsampled using scheme S. Furthermore, $\vec{x}_{\text{GT}}$ corresponds to the ground truth image (reconstructed from $\mathbf{y}$ using RSS), $\mathbf{x}_{\text{multi}}^{\text{S}}$ represents the image prediction obtained from the model trained on all subsampling schemes, and $\mathbf{x}_{\text{single}}^{\text{S}}$ signifies the image prediction derived from the model trained specifically on scheme S, with $\Tilde{\mathbf{y}}^{\text{S}}$ as input in both cases.

By computing the difference $m(\vec{x}_{\text{GT}}, \mathbf{x}_{\text{multi}}^{\text{S}}) - m(\vec{x}_{\text{GT}}, \mathbf{x}_{\text{single}}^{\text{S}})$ for each evaluation metric $m$, we can compare the performance of the two setups/models. Positive differences indicate superior performance for the multi-scheme model, particularly in the case of SSIM, pSNR, and SNR. Conversely, negative differences indicate better performance for the scheme-specific model, specifically for HFEN$_1$, HFEN$_2$, and NMSE. To assess the statistical significance, we conducted pair-wise ASO significance tests between the multi-scheme and scheme-specific quantitative results for each combination of dataset, acceleration, and metric.

\newpage

\begin{figure*}[!hbt]
    \centering
    \includegraphics[width=\textwidth]{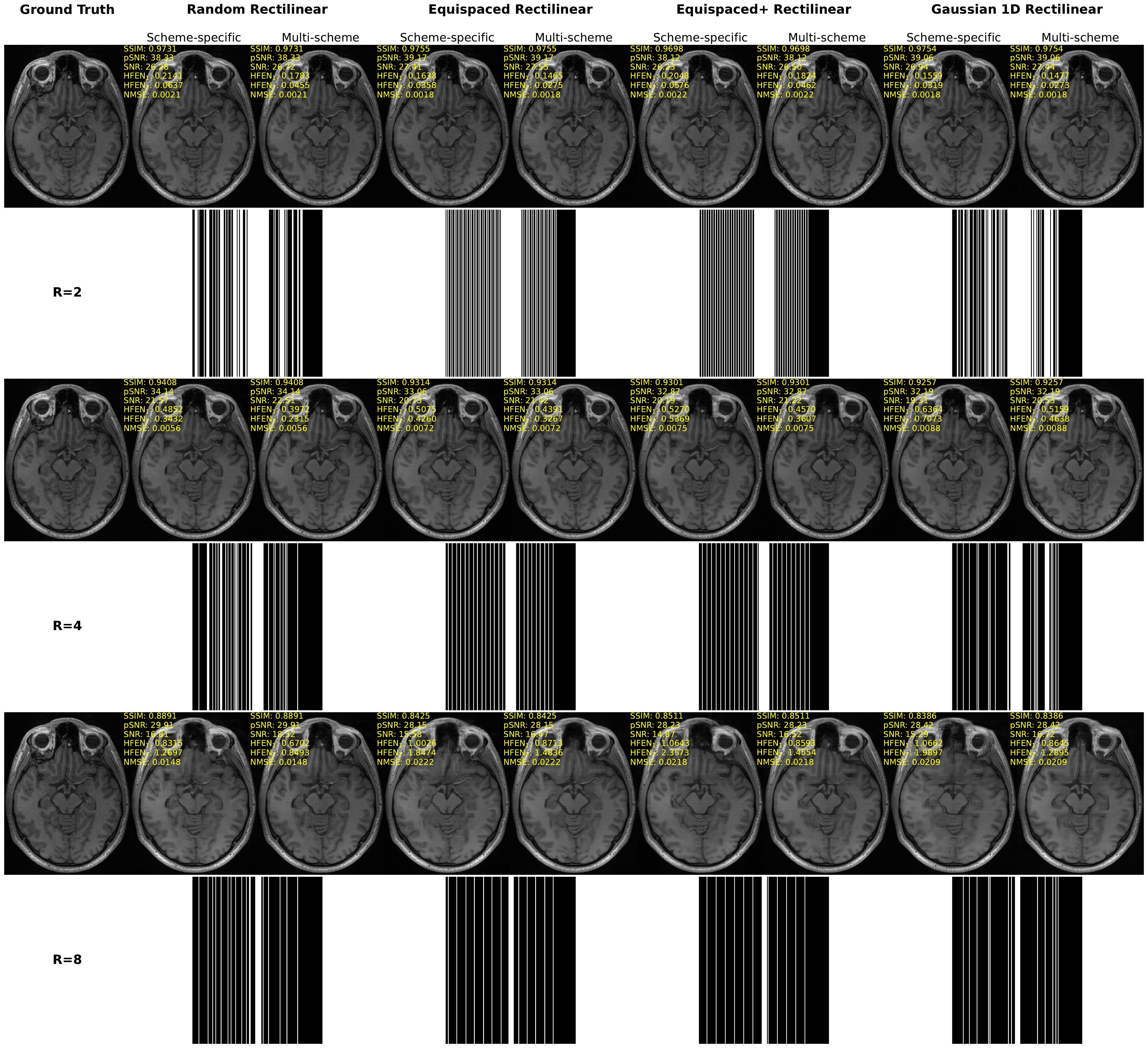}
    \caption{Scheme-specific vs multi-scheme setups visual comparison. Representative reconstructions of a sample from the Calgary-Campinas dataset subsampled with the four rectilinear schemes obtained from the scheme-specific and multi-scheme experiments for three acceleration factors.}
     \label{fig:A3}
\end{figure*}
\newpage

\subsection{Robustness to Model Choice}
\begin{figure*}[!hbt]
    \centering
    \includegraphics[width=\textwidth]{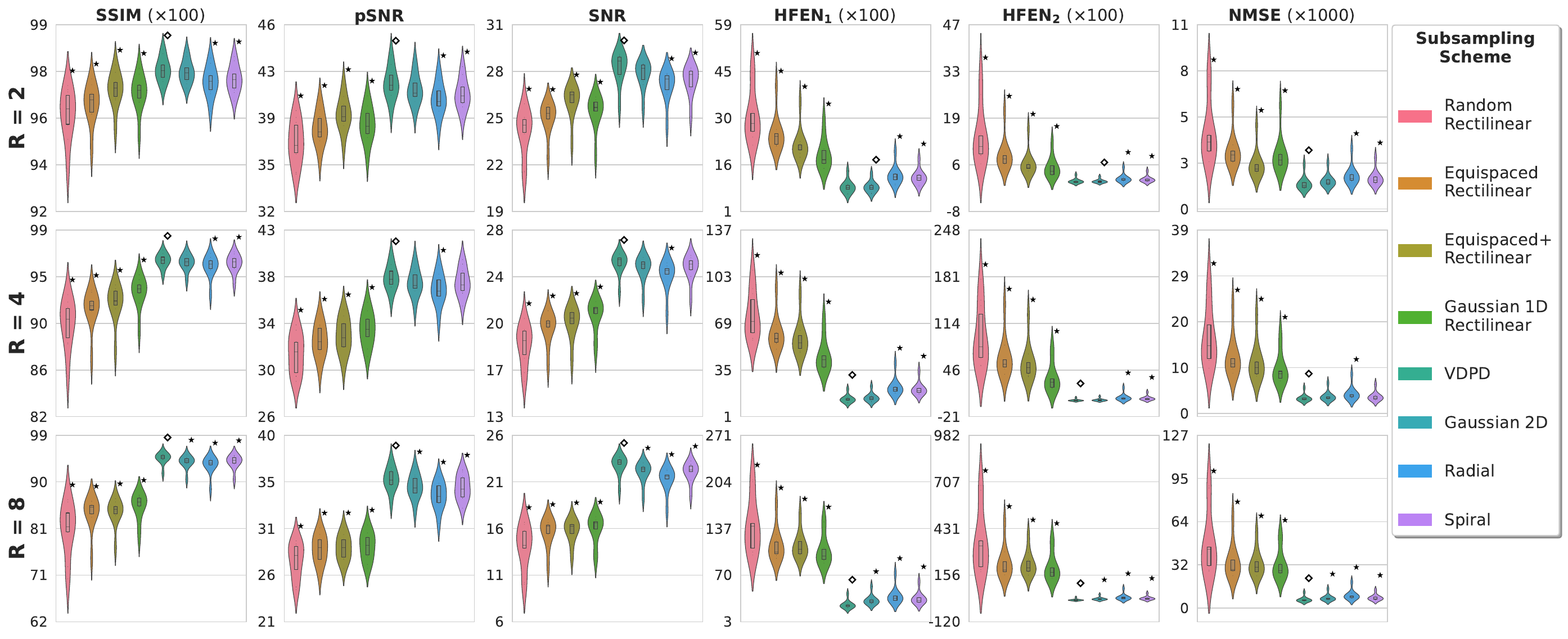}
    \caption{Model dependence experiments quantitative results on the Calgary-Campinas dataset using RIMs instead of RecurrentVarNets. For each subsampling scheme a distinct RIM was trained (in total 8 models). All models were built with identical hyperparameters and were trained to convergence. In each case we picked to evaluate the best model based on the validation performance on the SSIM metric. For each acceleration-metric combination, pair-wise ASO significance tests were performed between the average best (VDPD, denoted by $\diamond$) performing and the rest schemes. The symbol $\star$ indicates that VDPD was found to be significantly better ($\epsilon_\text{min} < 0.5$).}
     \label{fig:A4}
\end{figure*}

In this appendix subsection, we present in \Figure{A4} violin plots depicting the result metrics of our experiments, accompanied by \Table{A4} showing the average metrics. These experiments were conducted using RIMs as the deep learning architecture on the Calgary-Campinas dataset. Each subsampling scheme was trained with a separate RIM model, resulting in a total of 8 models. All models were trained until convergence with the same set of hyperparameters. 
\setlength{\tabcolsep}{0.8pt}
{\renewcommand{\arraystretch}{1.7}
\begin{table*}[!hbt]
 \centering
\caption{Model dependence experiments average quantitative results on the Calgary-Campinas dataset using RIMs instead of RecurrentVarNets.  Bold numbers indicate the best across subsampling schemes.}
\resizebox{\textwidth}{!}{%
\begin{tabular}{ccccccccccccccccccc}
\hline
\multirow{3}{*}{\begin{tabular}[c]{@{}c@{}}Subsampling\\ Scheme\end{tabular}} & \multicolumn{18}{c}{Acceleration Factor ($R$)}                                                                                                                                                                                                                                 \\ \cline{2-19} 
                                                                              & \multicolumn{6}{c}{2}                                                  & \multicolumn{6}{c}{4}                                                                             & \multicolumn{6}{c}{8}                                                                             \\ \cline{2-19} 
                                                                              & SSIM  & pSNR & SNR  & HFEN$_{1}$   & HFEN$_{2}$   & NMSE           & SSIM           & pSNR          & SNR           & HFEN$_{1}$   & HFEN$_{2}$   & NMSE           & SSIM           & pSNR          & SNR           & HFEN$_{1}$   & HFEN$_{2}$   & NMSE           \\ \hline
Random Rect.                                                                  & 0.959 & 37.2 & 24.0 & 0.309          & 0.143          & 0.005          & 0.903          & 31.7          & 18.6          & 0.748          & 0.945          & 0.016          & 0.805          & 27.2          & 13.8          & 1.403          & 3.493          & 0.047          \\
Equispaced Rect.                                                              & 0.963 & 38.3 & 24.9 & 0.255          & 0.088          & 0.003          & 0.917          & 32.8          & 19.9          & 0.618          & 0.646          & 0.012          & 0.836          & 28.3          & 15.3          & 1.138          & 2.297          & 0.034          \\
Equispaced+ Rect.                                                             & 0.969 & 39.4 & 26.0 & 0.221          & 0.063          & 0.003          & 0.922          & 33.3          & 20.3          & 0.583          & 0.566          & 0.011          & 0.838          & 28.5          & 15.4          & 1.127          & 2.231          & 0.033          \\
Gaussian  1D                                                                  & 0.968 & 38.6 & 25.3 & 0.191          & 0.051          & 0.003          & 0.932          & 33.9          & 20.9          & 0.457          & 0.351          & 0.010          & 0.849          & 28.6          & 15.6          & 1.063          & 2.031          & 0.032          \\
VDPD                                                                          & \textbf{0.977} & \textbf{41.8} & \textbf{28.2} & 0.091          & 0.011          & \textbf{0.002} & \textbf{0.962} & \textbf{38.4} & \textbf{24.9} & \textbf{0.137} & \textbf{0.023} & \textbf{0.003} & \textbf{0.943} & \textbf{35.7} & \textbf{22.3} & \textbf{0.276} & \textbf{0.096} & \textbf{0.006} \\
Gaussian  2D                                                                  & 0.976 & 41.3 & 27.8 & \textbf{0.090} & \textbf{0.010} & 0.002          & 0.960          & 38.0          & 24.6          & 0.147          & 0.028          & 0.004          & 0.936          & 34.9          & 21.6          & 0.342          & 0.159          & 0.008          \\
Radial                                                                        & 0.972 & 40.6 & 27.0 & 0.126          & 0.018          & 0.002          & 0.957          & 37.4          & 24.0          & 0.223          & 0.064          & 0.004          & 0.930          & 33.9          & 20.8          & 0.402          & 0.255          & 0.010          \\
Spiral                                                                        & 0.973 & 41.0 & 27.4 & 0.120          & 0.016          & 0.002          & 0.959          & 38.0          & 24.6          & 0.210          & 0.052          & 0.004          & 0.936          & 34.8          & 21.6          & 0.367          & 0.195          & 0.008          \\ \hline
\end{tabular}%
}
 \label{tab:A4}
\end{table*}
} % arraystrech 

\end{document}